\documentclass{ws-ijbc}

\usepackage{amsmath}
\usepackage{amssymb}
\usepackage{amsfonts}
\usepackage{verbatim}
\usepackage{graphicx}
\usepackage{palatino,url,multicol}
\usepackage{cite}

\usepackage{float}

\begin{document}

\catchline{}{}{}{}{} 

\markboth{}{}

\title{Bifurcations and Amplitude Death from Distributed Delays in Coupled Landau-Stuart Oscillators and a Chaotic Parametrically Forced van der Pol-Rayleigh System}

\date{\today }

\author{S. ROY CHOUDHURY}

\address{Department of Mathematics, University of Central Florida, Address\\
Orlando, Florida 32816, USA\\
choudhur@cs.ucf.edu}

\author{Ryan Roopnarain}

\address{Department of Mathematics, University of Central Florida, Address\\
Orlando, Florida 32816, USA\\
rdr0071772@knights.ucf.edu}

\maketitle

\begin{history}
\received{(to be inserted by publisher)}
\end{history}

\begin{abstract}
Distributed delays modeled by 'weak generic kernels' are introduced in the well-known coupled Landau-Stuart system, as well as a chaotic van der Pol-Rayleigh system with parametric forcing. The systems are closed  via the 'linear chain trick'. Linear stability analysis of the systems and conditions for Hopf bifurcation which initiates oscillations are investigated, including deriving the normal form at bifurcation, and deducing the stability of the resulting limit cycle attractor. The value of the delay parameter $a = a_{Hopf}$ at Hopf bifurcation picks out the onset of Amplitude Death(AD) in all three systems, with oscillations at larger values (corresponding to weaker delay).
In the Landau-Stuart system, the Hopf-generated limit cycles for $a > a_{Hopf}$  turn out to be remarkably stable under very large variations of all other system parameters beyond the Hopf bifurcation point, and do not undergo
further symmetry breaking, cyclic-fold, flip, transcritical or Neimark-Sacker bifurcations. This is to be expected as the corresponding undelayed systems are robust oscillators over very wide ranges of their respective parameters. Numerical simulations reveal strong distortion and rotation of the limit cycles in phase space as the parameters are pushed far into the post-Hopf regime, and also reveal other features, such as how the oscillation amplitudes and time periods of the physical variables on the limit cycle attractor change as the delay and other parameters are varied. For the chaotic system, very strong delays may still lead to the cessation of oscillations and the onset of AD (even for relatively large values of the system forcing which tends to oppose this phenomenon). Varying of the other important system parameter, the parametric excitation, leads to a rich sequence of dynamical behaviors, with the bifurcations leading from one regime (or type of attractor) into the next being carefully tracked.

\end{abstract}

\keywords{amplitude death; distributed delays; bifurcation analysis; chaotic attractor}

\section{Introduction}

As is well-known, nonlinear dynamical systems, especially coupled ones, are of wide interest in many areas of science and technology. When such systems which, in isolation are capable of a great variety of behaviors, are coupled, a host of novel phenomena are seen. These depend on the specific features, both of the individual systems, as well as the type of coupling.

One important area of application of such systems is what might imprecisely be referred to as 'stabilization', i.e., the creation of simpler system attractors via the coupling. The best known among these is suppression of oscillations, most often termed as Amplitude Death (AD)\citep{SPR} , even when the uncoupled systems themselves do not exhibit such stationary behavior. Coupling-induced AD is an instance of a more general phenomenon that may include actual cessation of oscillations, or the conversion of chaotic dynamics to periodic or quasiperiodic dynamics. In the case of oscillation suppression by coupling, two separate phenomena are now recognized. The first is suppression of oscillation to a single or homogeneous steady state (nowadays referred to as AD), versus the second or Oscillation Death (OD)\citep{KVK} where the oscillators asymptotically populate different fixed points or 'inhomogeneous steady states', some of which may not have been stable, or perhaps not even present, for the uncoupled oscillators.

Both AD and OD are known to occur in various settings. These are reviewed in \citep{SPR} and \citep{KVK}, and include mismatched oscillators\citep{CF}, \citep{BE}, \citep{KBE}, and \citep{AEK}, delayed interactions \citep{RSJ},\citep{RSJ2},\citep{RSJ3},\citep{RSJ4}, and \citep{LS} (including distributed delays\citep{A} and cumulative signals\citep{SPR1} and \citep{SPR2}), conjugate coupling\citep{KR1}, \citep{KR2}, \citep{KPPR}, \citep{KRP}, and \citep{ZWP}, dynamic coupling \citep{KK}, nonlinear coupling\citep{PDAR} and \citep{PLGK}, linear augmentation \citep{SSSP} and \citep{RAA},velocity coupling \citep{SPR}, and other schemes.

In this paper, we consider the effect of distributed delays on a variety of coupled systems carefully. While discrete delays have been considered in some detail, distributed delay effects are less-investigated, although they are known to provide stronger AD or OD effects. In order to facilitate analytical investigation to the extent possible, we use the so-called 'chain trick' together with the 'weak generic kernel' form of distributed delay\citep{Choud}, \citep{Cush}, and \citep{Mac}. We consider the effect of incorporating such delays in two different models viz. two different Van der Pol type oscillators, and a chaotic oscillator\citep{Warm}.

The remainder of this paper is organized as follows. Section 2 briefly reviews the linear stability analysis of the two oscillator systems above in the absence of delay, while Section 3 repeats that analysis with the inclusion of 'weak generic kernel' delays in some of the nonlinear interaction terms, thus giving a first set of modifications of the dynamics. The normal form at Hopf bifurcation is derived in Section 4. Section 5 then considers detailed numerical results contrasting the behavior of the undelayed systems to the modifications created by the weak generic delays. Finally, Section 6 summarizes the results and conclusions.

\section{Linear Stability}

In this section we briefly recapitulate the linear stability of the undelayed systems we will be considering. 


\subsection{The Landau-Stuart Equation}
The coupled Landau-Stuart system is given by \citep{RSJ},\citep{RSJ2},\citep{RSJ3},\citep{RSJ4}, and \citep{LS}
\begin{align}
\dot{z}_1(t) &= ( 1 + i \omega_1 - |z_1(t)|^2)z_1(t) + \varepsilon(z_2(t) - z_1(t))\notag \\
\dot{z}_2(t) &= (1 +i\omega_2 - |z_2(t)|^2)z_2(t) + \varepsilon(z_1(t)-z_2(t))
\end{align}
where $z_i(t)$ are complex and $\omega_i>0$ for $i=1,2$ and $\varepsilon>0$. In order to work with the system we first convert it in to a real system by defining $z_k(t) = x_k(t) + i y_k(t)$ for each $k =1,2$ which gives:
\begin{align}
\label{undelaysystem}
\dot{x}_1 & = x_1 - \omega_1 y_1 - (x_1^2 + y_1^2)x_1 + \varepsilon(x_2 - x_1)\notag\\
\dot{y}_1 &= y_1 + \omega_1 x_1 - (x_1^2 + y_1^2)y_1 + \varepsilon(y_2 - y_1)\notag\\
\dot{x}_2 &= x_2 - \omega_2 y_2 - (x_2^2 + y_2^2)x_2 + \varepsilon(x_1 - x_2)\notag\\
\dot{y}_2 &= y_2 + \omega_2 x_2 - (x_2^2 + y_2^2)y_2 + \varepsilon(y_1 - y_2)\notag\\
\end{align}
 The only fixed point of this system is the trivial one $P$: 
\begin{equation}
P = (x_{1,0}, y_{1,0}, x_{2,0}, y_{2,0}) = (0, 0, 0, 0)
\end{equation}
The Jacobian matrix of \eqref{undelaysystem} is given by:
\begin{equation}\label{undelayjacob1}
\begin{pmatrix}
1-\varepsilon-3x_1^2-y_1^2  &   -\omega_1 - 2x_1y_1 & \varepsilon   & 0\\
\omega_1 -2x_1y_1   & 1-\varepsilon-x_1^2-3y_1^2 & 0 & \varepsilon\\
\varepsilon & 0 & 1-\varepsilon-2x_2^2 - y_2^2 & -\omega_2 -2x_2y_2\\
0 & \varepsilon & \omega_2 - 2x_2y_2 & 1-\varepsilon-x_2^2-3y_2^2
\end{pmatrix}
\end{equation}
and evaluating at the fixed point $P$ gives:
\begin{equation}
    \label{eundelayedjacob1}
    \begin{pmatrix}
    1-\varepsilon & -\omega_1 & \varepsilon & 0\\
    \omega_1 & 1-\varepsilon & 0 & \varepsilon\\
    \varepsilon & 0 & 1-\varepsilon & -\omega_2\\
    0 & \varepsilon & \omega_2 & 1-\varepsilon
    \end{pmatrix}
\end{equation}
The eigenvalues of this matrix then satisfy the characteristic equation (to be considered later)
\begin{align}
   %
 %
  %
     &\lambda^4+(-4 + 4 \varepsilon) \lambda^3 + (6 - 12 \varepsilon + 4 \varepsilon^2 + \omega_1^2 + \omega_2^2) \lambda^2 \notag\\
     &\quad + (-4 + 12 \varepsilon - 8 \varepsilon^2 - 2 \omega_1^2 + 2 \varepsilon \omega_1^2 - 2 \omega_2^2 + 2 \varepsilon \omega_2^2)\lambda\notag\\
     &\quad + (1 - 4 \varepsilon + 4 \varepsilon^2 + \omega_1^2 - 2 \varepsilon \omega_1^2 + \varepsilon^2 \omega_1^2 + 2 \varepsilon^2 \omega_1 \omega_2 \notag\\
     &\quad\quad + \omega_2^2 - 2 \varepsilon \omega_2^2 + \varepsilon^2 \omega_2^2 +\omega_1^2 \omega_2^2  ) = 0
\end{align}

\noindent
which will be considered later.


\subsection{Chaotic System}
The chaotic system we consider is a coupled van der Pol-Rayleigh oscillator system with parametric excitation, and is given by\citep{Warm}
\begin{align}
\ddot{x} + (-\alpha_1 + \beta_1\dot{x}^2)\dot{x} + \delta_1 x + \gamma_1 x^3 + (\delta_{12} - \mu\cos(2\nu t))(x-y)&=q\cos(\nu t)\notag \\
\ddot{y} + M(-\alpha_2 + \beta_2\dot{y}^2)\dot{y} + M\delta_2 y + \gamma_2 y^3  -M(\delta_{12} - \mu\cos(2\nu t))(y-x)&=0
\end{align}
In order to work with the system we first convert it in to a first-order system by defining $x_1(t) = x(t), x_2(t) = \dot{x}(t), y_1(t) = y(t), y_2(t) = \dot{y}(t)$ which gives:
\begin{align}
\label{undelaysystem1}
\dot{x}_1 & = x_2 \notag\\
\dot{x}_2 &= (\alpha_1 - \beta_1x_2^2)x_2 - \delta_1 x_1 - \gamma_1 x_1^3 - (\delta_{12} - \mu\cos(2\nu t))(x_1-y_1)+q\cos(\nu t)\notag\\
\dot{y}_1 &= y_2\notag\\
\dot{y}_2 &= M(\alpha_2 - \beta_2y_2^2)y_2  - M\delta_2 y_1 - \gamma_2 y_1^3  +M(\delta_{12} - \mu\cos(2\nu t))(y_1-x_1)
\end{align}
Considering the homogeneous system $q=0$, we find the fixed point: 
\begin{equation}
P_0 = (x_{1,0}, x_{2,0}, y_{1,0},y_{2,0}) = (0, 0, 0, 0)
\end{equation}
and if, in addition, we have $\delta_1/\gamma_1 = \delta_2/\gamma_2 <0$ then there are two additional fixed points:
\begin{align}
    P_1 = (x_{1,1}, x_{2,1}, y_{1,1},y_{2,1}) &= \left(\sqrt{-\frac{\delta_1}{\gamma_1}}, 0, \sqrt{-\frac{\delta_1}{\gamma_1}}, 0\right)\\
    P_2 = (x_{1,2}, x_{2,2}, y_{1,2},y_{2,2}) &= \left(-\sqrt{-\frac{\delta_1}{\gamma_1}}, 0, -\sqrt{-\frac{\delta_1}{\gamma_1}}, 0\right)
\end{align}
Next we convert the system to an autonomous system by defining $T(t) = t$:
\begin{align}
\label{c1auto}
\dot{T} &= 1\notag\\
\dot{x}_1 & = x_2 \notag\\
\dot{x}_2 &= (\alpha_1 - \beta_1x_2^2)x_2 - \delta_1 x_1 - \gamma_1 x_1^3 - (\delta_{12} - \mu\cos(2\nu T))(x_1-y1)+q\cos(\nu T)\notag\\
\dot{y}_1 &= y_2\notag\\
\dot{y}_2 &= M(\alpha_2 - \beta_2y_2^2)y_2  - M\delta_2 y_1 - \gamma_2 y_1^3  +M(\delta_{12} - \mu\cos(2\nu T))(y_1-x_1)
\end{align}
The Jacobian matrix of \eqref{c1auto} is given by:
\begin{equation}\label{undelayjacob}
\begin{pmatrix}
0&0&0&0&0\\
0&0  &  1 & 0   & 0\\
c_1&c_2  & \alpha_1 - 3\beta_1x_2^2 & \delta_{12} - \mu\cos(2\nu T) & 0\\
0&0 & 0 & 0 & 1\\
c_3&M(\delta_{12} - \mu\cos(2\nu T) & 0 &c_4 & M(\alpha_2 -3\beta_2y_2^2)
\end{pmatrix}
\end{equation}
where
\begin{align}
    c_1 &= -2\mu\nu (x_1-y_1)\sin(2\nu T)\\
    c_2 &= -\delta_1-\delta_{12} - 3\gamma_1x_1^2+\mu\cos(2\nu T)\\
    c_3 &= 2M\mu\nu (x_1-y_1)\sin(2\nu T)\\
    c_4 &=  M(-\delta_2 - 3\gamma_2y_1^2 -\delta_{12} + \mu\cos(2\nu T))
\end{align}
and evaluating at the fixed point $P_0$ gives:
\begin{equation}
    \label{eundelayedjacob}
    \begin{pmatrix}
    0&0&0&0&0\\
    0&0  &  1 & 0   & 0\\
    0&-\delta_1-\delta_{12} +\mu\cos(2\nu T)  & \alpha_1  & \delta_{12} - \mu\cos(2\nu T) & 0\\
    0&0 & 0 & 0 & 1\\
    0&M(\delta_{12} - \mu\cos(2\nu T) & 0 & M(-\delta_2 -\delta_{12} + \mu\cos(2\nu T))  & M(\alpha_2 -3\beta_2y_2^2)
\end{pmatrix}
\end{equation}
The eigenvalues of this matrix then satisfy the characteristic equation which will be considered later
\begin{align}
     &\lambda(\lambda^4+(-\alpha_1-\alpha_2 M) \lambda^3 + (\delta_1+\delta_{12}+\alpha_1 \alpha_2 M-\mu \cos(2\nu T)\notag\\
     &\quad +M (\delta_{12}+\delta_2-\mu \cos(2\nu T))) \lambda^2 +(-\alpha_2 \delta_1 M-\alpha_2 \delta_{12} M+\alpha_2 M \mu \cos(2\nu T)\notag\\
     &\quad-\alpha_1 M (\delta_{12}+\delta_2-\mu \cos(2\nu T)))\lambda + -M (\delta_{12}-\mu \cos(2\nu T))^2\notag\\
     &\quad+\delta_1 M (\delta_{12}+\delta_2-\mu \cos(2\nu T))+\delta_{12} M (\delta_{12}+\delta_2-\mu \cos(2\nu T))\notag\\
     &\quad-M \mu \cos(2\nu T) (\delta_{12}+\delta_2-\mu \cos(2\nu T))) = 0
\end{align}
Next, evaluating the Jacobian at either of the fixed points $P_1$ or $P_2$ gives the matrix:
\begin{equation}
    \label{eundelayedjacob2}
    \begin{pmatrix}
    0&0&0&0&0\\
    0&0  &  1 & 0   & 0\\
    0&2\delta_1-\delta_{12} +\mu\cos(2\nu T)  & \alpha_1  & \delta_{12} - \mu\cos(2\nu T) & 0\\
    0&0 & 0 & 0 & 1\\
    0&M(\delta_{12} - \mu\cos(2\nu T) & 0 & M(2\delta_2 -\delta_{12} + \mu\cos(2\nu T))  & M\alpha_2
\end{pmatrix}
\end{equation}
The eigenvalues of this matrix then satisfy the characteristic equation (to be considered later):
\begin{align}
   &\lambda(\lambda^4 + (-\alpha_1-\alpha_2 M)\lambda^3 + (-2 \delta_1+\delta_{12}+\alpha_1 \alpha_2 M-\mu \cos(2\nu T)-M (-\delta_{12}+2 \delta_2\notag\\
   &\quad+\mu \cos(2\nu T)))\lambda^2 + (2 \alpha_2 \delta_1 M-\alpha_2 \delta_{12} M+\alpha_2 M \mu \cos(2\nu T)+\alpha_1 M (-\delta_{12}+2 \delta_2\notag\\
   &\quad +\mu \cos(2\nu T)))\lambda -M (\delta_{12}-\mu \cos(2\nu T))^2+2 \delta_1 M (-\delta_{12}+2 \delta_2+\mu \cos(2\nu T))\notag\\
   &\quad-\delta_{12} M (-\delta_{12}+2 \delta_2+\mu \cos(2\nu T))+M \mu \cos(2\nu T) (-\delta_{12}+2 \delta_2+\mu \cos(2\nu T))) =0
\end{align}



\section{Linear Stability and Hopf Bifurcation Analysis of the Delayed Systems}

In this section we introduce the delayed systems and perform the linear stability and Hopf bifurcation analysis on them. 

\subsection{Delayed Landau-Stuart Equation}
Now we consider here the case where the Landau-Stuart oscillators are coupled with a weak distributed time delay in the first equation:
\begin{align}
\label{cdelaysys}
\dot{z}_1(t) &= ( 1 + i \omega_1 - |z_1(t)|^2)z_1(t) + \varepsilon\left(\int_{-\infty}^t z_2(\tau) a e^{-a(t-\tau)}d\tau - z_1(t)\right)\notag \\
\dot{z}_2(t) &= (1 +i\omega_2 - |z_2(t)|^2)z_2(t) + \varepsilon(z_1(t)-z_2(t))
\end{align}
By defining 
\begin{equation}
z_3(t) = \int_{-\infty}^t z_2(\tau)ae^{-a(t-\tau)}d\tau\notag
\end{equation}
we can reduce the system \eqref{cdelaysys} to the system of differential equations:
\begin{align}
\dot{z}_1(t) &= ( 1 + i \omega_1 - |z_1(t)|^2)z_1(t) + \varepsilon(z_3(t) - z_1(t))\notag \\
\dot{z}_2(t) &= (1 +i\omega_2 - |z_2(t)|^2)z_2(t) + \varepsilon(z_1(t)-z_2(t)) \notag \\
\dot{z}_3(t) &= a(z_2 - z_3)
\end{align}
As in the undelayed case, in order to work with this system we convert it to a real system by defining $z_k(t) = x_k(t) + i y_k(t)$ for each $k =1,2,3$,  which gives:
\begin{align}
\label{lsdelay}
\dot{x}_1 & = x_1 - \omega_1 y_1 - (x_1^2 + y_1^2)x_1 + \varepsilon(x_3 - x_1)\notag\\
\dot{y}_1 &= y_1 + \omega_1 x_1 - (x_1^2 + y_1^2)y_1 + \varepsilon(y_3 - y_1)\notag\\
\dot{x}_2 &= x_2 - \omega_2 y_2 - (x_2^2 + y_2^2)x_2 + \varepsilon(x_1 - x_2)\notag\\
\dot{y}_2 &= y_2 + \omega_2 x_2 - (x_2^2 + y_2^2)y_2 + \varepsilon(y_1 - y_2)\notag\\
\dot{x}_3 &= a(x_2-x_3)\notag\\
\dot{y}_3 &= a(y_2-y_3)
\end{align}

The only fixed point of this system is the trivial one $P$: 
\begin{equation}
P = (x_{1,0}, y_{1,0}, x_{2,0}, y_{2,0}, x_{3,0},y_{3,0}) = (0, 0, 0, 0, 0, 0)
\end{equation}

The Jacobian matrix of \eqref{lsdelay} is:

\begin{equation}
\label{jacob1}
\begin{pmatrix} 
1-\varepsilon - 3x_{1}^2 - y_{1}^2 & -\omega_1 - 2x_{1}y_{1} & 0 & 0 & \varepsilon & 0 \\
\omega_1 - 2x_{1}y_{1} & 1-\varepsilon -x_{1}^2 - 3y_{1}^2 & 0 & 0 & 0 & \varepsilon\\
\varepsilon & 0 & 1-\varepsilon - 3x_{2}^2 - y_{2}^2 & -\omega_2 - 2x_{2}y_{2} & 0 & 0\\
0 & \varepsilon &  \omega_2 - 2 x_{2}y_{2} & 1-\varepsilon - x_{2}^2 - 3y_{2}^2 & 0 & 0\\
0 & 0 & a & 0 & -a & 0\\
0 & 0 & 0 & a & 0  & -a
\end{pmatrix}
\end{equation}
which, evaluated at the fixed point $P$, gives:
\begin{equation}
\label{jacob2}
\begin{pmatrix}
1-\varepsilon & -\omega_1  & 0 & 0 & \varepsilon & 0 \\
\omega_1& 1-\varepsilon & 0 & 0 & 0 & \varepsilon\\
\varepsilon & 0 & 1-\varepsilon & -\omega_2  & 0 & 0\\
0 & \varepsilon &  \omega_2  & 1-\varepsilon & 0 & 0\\
0 & 0 & a & 0 & -a & 0\\
0 & 0 & 0 & a & 0  & -a
\end{pmatrix}
\end{equation}
The eigenvalues of this matrix satisfy the characteristic equation
\begin{equation} 
\label{efunc}
\lambda^6+ b_1 \lambda^5 + b_2 \lambda^4 + b_3 \lambda^3 +b_4\lambda^2 + b_5\lambda + b_6 =0
\end{equation}
where\\
\begin{align}
b_1 &=  2 (-2 + a + 2 \varepsilon) \notag \\
b_2 &=  6 + a^2 + 8 a (-1 + \varepsilon) - 12 \varepsilon + 6 \varepsilon^2 + \omega_1^2 + \omega_2^2\notag \\
b_3 &= 2 (2 a^2 (-1 + \varepsilon) + (-1 + \varepsilon) (2 - 4 \varepsilon + 2 \varepsilon^2 + \omega_1^2 + 
      \omega_2^2) \notag\\
&\quad + a (6 - 12 \varepsilon + 5 \varepsilon^2 + \omega_1^2 + \omega_2^2)) \notag \\
b_4 &= (1 - 2 \varepsilon + \varepsilon^2 + \omega_1^2) (1 - 2 \varepsilon + \varepsilon^2 + \omega_2^2) + 4 a (-1 + \varepsilon) (2 - 4 \varepsilon + \varepsilon^2 + \omega_1^2 + \omega_2^2) \notag\\
&\quad + a^2 (6 - 12 \varepsilon + 4 \varepsilon^2 + \omega_1^2 + \omega_2^2) \notag \\
b_5 &= 2 a (-2 \varepsilon^3 + (1 + \omega_1^2) (1 + \omega_2^2) - 2 \varepsilon (2 + \omega_1^2 + \omega_2^2) + \varepsilon^2 (5 + \omega_1^2 + \omega_1 \omega_2 + \omega_2^2) \notag \\
&\quad -a (2 + 4 \varepsilon^2 + \omega_1^2 + \omega_2^2 - \varepsilon (6 + \omega_1^2 + \omega_2^2))) \notag \\
b_6 &= a^2 ((1 + \omega_1^2) (1 + \omega_2^2) - 2 \varepsilon (2 + \omega_1^2 + \omega_2^2) + \varepsilon^2 (4 + \omega_1^2 + 2 \omega_1 \omega_2 + \omega_2^2)) 
\end{align}
\\
For $P$ to be a stable fixed point within the linearized analysis, all the eigenvalues must have negative real parts. From the Routh-Hurwitz criteria, the necessary and sufficient conditions for \eqref{lsdelay} to have $\hbox{Re} (\lambda_{1,2,3,4,5,6} <0)$ are:
\begin{align}
\label{rh1}
b_1 &>0 \\ 
\label{rh6}
b_6 &>0 \\ 
\label{rh2}
b_1 b_2 - b_3 &>0 \\ 
\label{rh3}
b_1 (b_2 b_3 + b_5) -b_3^2 - b_1^2 b_4 &>0 \\
\label{rh4}
b_1 (b_2 b_3 b_4 - b_2^2 b_5 + 2 b_4 b_5 - b_3 b_6)-b_3^2 b_4- b_5^2 \quad &\notag \\
 +b_2 b_3 b_5  + b_1^2 (-b_4^2 + b_2 b_6) &>0 \\  
  \label{hopfcondition}
 -b_3^2 b_4 b_5 + b_2 b_3 b_5^2 - b_5^3 + b_3^3 b_6 - b_1^3 b_6^2 + b_1^2 (-b_4^2 b_5 + b_3 b_4 b_6 + 2 b_2 b_5 b_6)\quad & \notag \\ 
 - b_1 (b_2^2 b_5^2 + b_2 b_3 (-b_4 b_5 + b_3 b_6) + b_5 (-2 b_4 b_5 + 3 b_3 b_6)) &>0 
\end{align}

When the final condition \eqref{hopfcondition} becomes an equality, the characteristic polynomial has one pair of purely imaginary complex conjugate roots. Here, we consider $a$ to be the bifurcation parameter and denote the left hand side of $\eqref{hopfcondition}$ by $f(a)$ which is a ninth degree polynomial in $a$ whose coefficients, which are too large to include, depend on $\omega_1, \omega_2,$ and $\varepsilon$. In order to solve the above conditions for parameter sets possibly leading to a Hopf bifurcation, we must first fix a value for $\varepsilon$. Then, with our fixed value of $\varepsilon$, we reduce the conditions \eqref{rh1} to \eqref{rh4} along with the condition $f(a) =0$ using computer algebra, to obtain conditions on the remaining parameters that may possibly lead to a Hopf bifurcation in the delayed system. 

For example, fixing $\varepsilon = 2$, one of the several sets of conditions for a Hopf bifurcation we obtain is that $$0<\omega_2\leq \sqrt{3}$$
$$
-\frac{2\omega_2}{1-\omega_2^2} + \sqrt{\frac{3+6\omega_2^2 - \omega_2^4}{(1+\omega_2^2)^2}} < \omega_1 <  \omega_2+2\sqrt{3}
$$
and that $a$ is the second root\footnote{when the roots are ordered in increasing real part, with real roots listed before complex roots and complex conjugate pairs listed next to each other} of the polynomial:
\begin{align*}
&(-12 + \omega_1^2 +  \omega_1^4 + 22  \omega_1  \omega_2 + 2  \omega_1^3  \omega_2 +  \omega_2^2 - 
    2  \omega_1^2  \omega_2^2 +  \omega_1^4  \omega_2^2 + 2  \omega_1  \omega_2^3 - 2  \omega_1^3  \omega_2^3 + 
     \omega_2^4 +  \omega_1^2  \omega_2^4)\\
     &\quad + (-96 - 16  \omega_1^2 + 2  \omega_1^4 + 32  \omega_1  \omega_2 - 
       8  \omega_1^3  \omega_2 - 16  \omega_2^2 + 12  \omega_1^2  \omega_2^2 - 8  \omega_1  \omega_2^3 + 
       2  \omega_2^4) x \\
       &\quad+ (-216 + 6  \omega_1^2 +  \omega_1^4 - 36  \omega_1  \omega_2 - 
       2  \omega_1^3  \omega_2 + 6  \omega_2^2 + 2  \omega_1^2  \omega_2^2 - 2  \omega_1  \omega_2^3 + 
        \omega_2^4) x^2\\
        &\quad+ (-96 + 8  \omega_1^2 - 16  \omega_1  \omega_2 + 
       8  \omega_2^2) x^3 + (-12 +  \omega_1^2 - 2  \omega_1  \omega_2 +  \omega_2^2) x^4 
\end{align*}
In particular we can fix $\omega_2 = 15$ to obtain the condition $11.5359<\omega_1<18.4641$. Then fixing $\omega_1$, say to $\omega_1 = 15$, we obtain that $a$ be the second root of the polynomial $-12(-17324 + 8x + 468x^2 +8x^3 + x^4)$, or $a\approx 5.63185$. So we have that the parameter set $(\varepsilon, \omega_1,\omega_2,a) = (2,15,15,5.63185)$ possibly results in a Hopf bifurcation in the delayed system.


\subsection{Delayed Chaotic System }
Now we consider here the case where the Landau-Stuart oscillators are coupled with a weak distributed time delay in the first equation:
\begin{align}
\label{cdelay}
\ddot{x} + (-\alpha_1 + \beta_1\dot{x}^2)\dot{x} + \delta_1 x + \gamma_1 x^3 + (\delta_{12} - \mu\cos(2\nu t))(x-z)&=q\cos(\nu t)\notag \\
\ddot{y} + M(-\alpha_2 + \beta_2\dot{y}^2)\dot{y} + M\delta_2 y + \gamma_2 y^3  -M(\delta_{12} - \mu\cos(2\nu t))(y-x)&=0
\end{align}
where  
\begin{equation}
z(t) = \int_{-\infty}^t y(\tau)ae^{-a(t-\tau)}d\tau\notag
\end{equation}
and we can reduce the system \eqref{cdelay} to the system of differential equations:
\begin{align}
\ddot{x} + (-\alpha_1 + \beta_1\dot{x}^2)\dot{x} + \delta_1 x + \gamma_1 x^3 + (\delta_{12} - \mu\cos(2\nu t))(x-z)&=q\cos(\nu t)\notag \\
\ddot{y} + M(-\alpha_2 + \beta_2\dot{y}^2)\dot{y} + M\delta_2 y + \gamma_2 y^3  -M(\delta_{12} - \mu\cos(2\nu t))(y-x)&=0\notag\\
\dot{z} -a(y-z) &=0
\end{align}
As in the undelayed case, we first convert it in to a first-order system by defining $x_1(t) = x(t), x_2(t) = \dot{x}(t), y_1(t) = y(t), y_2(t) = \dot{y}(t)$ which gives:
\begin{align}
\label{c1delaysys}
\dot{x}_1 & = x_2 \notag\\
\dot{x}_2 &= (\alpha_1 - \beta_1x_2^2)x_2 - \delta_1 x_1 - \gamma_1 x_1^3 - (\delta_{12} - \mu\cos(2\nu t))(x_1-z)+q\cos(\nu t)\notag\\
\dot{y}_1 &= y_2\notag\\
\dot{y}_2 &= M(\alpha_2 - \beta_2y_2^2)y_2  - M\delta_2 y_1 - \gamma_2 y_1^3  +M(\delta_{12} - \mu\cos(2\nu t))(y_1-x_1)\notag\\
\dot{z} &= a(y_1-z)
\end{align}

The fixed points of the delayed system are: 
\begin{equation}
P_0 = P_0 = (x_{1,0}, x_{2,0}, y_{1,0},y_{2,0}, z_0) = (0, 0, 0, 0,0)
\end{equation}
and if, in addition, we have $\delta_1/\gamma_1 = \delta_2/\gamma_2 <0$ then there are two additional fixed points:
\begin{align}
    P_1 = (x_{1,1}, x_{2,1}, y_{1,1},y_{2,1},z_1) &= \left(\sqrt{-\frac{\delta_1}{\gamma_1}}, 0, \sqrt{-\frac{\delta_1}{\gamma_1}}, 0,\sqrt{-\frac{\delta_1}{\gamma_1}}\right)\\
    P_2 = (x_{1,2}, x_{2,2}, y_{1,2},y_{2,2},z_2) &= \left(-\sqrt{-\frac{\delta_1}{\gamma_1}}, 0, -\sqrt{-\frac{\delta_1}{\gamma_1}}, 0,-\sqrt{-\frac{\delta_1}{\gamma_1}}\right)
\end{align}
However, in what follows the parameter regimes we will consider will include the case $\delta_1/\gamma_1 = \delta_2/\gamma_2$, and so these two additional fixed points will not exist in our case. Next we convert the system to an autonomous system by defining $T(t) = t$:
\begin{align}
\label{c1delayauto}
\dot{T} &= 1\notag\\
\dot{x}_1 & = x_2 \notag\\
\dot{x}_2 &= (\alpha_1 - \beta_1x_2^2)x_2 - \delta_1 x_1 - \gamma_1 x_1^3 - (\delta_{12} - \mu\cos(2\nu t))(x_1-z)+q\cos(\nu t)\notag\\
\dot{y}_1 &= y_2\notag\\
\dot{y}_2 &= M(\alpha_2 - \beta_2y_2^2)y_2  - M\delta_2 y_1 - \gamma_2 y_1^3  +M(\delta_{12} - \mu\cos(2\nu t))(y_1-x_1)\notag\\
\dot{z} &= a(y_1-z)
\end{align}
The Jacobian matrix of \eqref{c1delayauto} is:

\begin{equation}
\label{jacob11}
\begin{pmatrix}
0&0&0&0&0&0\\
0&0  &  1 & 0   & 0 & 0\\
c_1&c_2  & \alpha_1 - 3\beta_1x_2^2 & 0 & 0 &\delta_{12} - \mu\cos(2\nu T)\\
0&0 & 0 & 0 & 1 & 0\\
c_3&M(\delta_{12} - \mu\cos(2\nu T) & 0 &c_4 & M(\alpha_2 -3\beta_2y_2^2)&0\\
0&0 & 0 & a & 0 & -a
\end{pmatrix}
\end{equation}
where
\begin{align}
    c_1 &=-2\mu\nu(x_1-z)\sin(2\nu T)\\
    c_2 &= -\delta_1-\delta_{12} - 3\gamma_1x_1^2+\mu\cos(2\nu T)\\
    c_3 &= 2M\mu\nu(x_1-y_1)\sin(2\nu T)\\
    c_4 &=  M(-\delta_2 - 3\gamma_2y_1^2 -\delta_{12} + \mu\cos(2\nu T))
\end{align}
Evaluating at the fixed point $P_0$ of the original nonautonomous system gives:
\begin{equation}
\label{jacob21}
\begin{pmatrix}
0&0&0&0&0&0\\
0&0  &  1 & 0   & 0 & 0\\
0&-\delta_1-\delta_{12} +\mu\cos(2\nu T)  & \alpha_1  & 0 & 0 &\delta_{12} - \mu\cos(2\nu T)\\
0&0 & 0 & 0 & 1 & 0\\
0&M(\delta_{12} - \mu\cos(2\nu T) & 0 & M(-\delta_2 -\delta_{12} + \mu\cos(2\nu T)) & M\alpha_2 &0\\
0&0 & 0 & a & 0 & -a
\end{pmatrix}
\end{equation}
The eigenvalues of this matrix satisfy the characteristic equation
\begin{equation} 
\label{efunc1}
\lambda(\lambda^5 + b_1\lambda^4 + b_2 \lambda^3 +b_3\lambda^2 + b_4\lambda + b_5) =0
\end{equation}
where\\
\begin{align}
b_1 &=a-\alpha_1-\alpha_2 M \notag \\
b_2 &= -a \alpha_1-a \alpha_2 M+\alpha_1 \alpha_2 M+\delta_1+\delta_{12} M+\delta_{12}+\delta_2 M-M \mu \cos (2 \nu T)-\mu \cos (2 \nu T) \notag \\
b_3 &= a \alpha_1 \alpha_2 M+a \delta_1+a \delta_{12} M+a \delta_{12}+a \delta_2 M-a M \mu \cos (2 \nu T)-a \mu \cos (2 \nu T)\notag\\
&\quad-\alpha_1 \delta_{12} M-\alpha_1 \delta_2 M+\alpha_1 M \mu \cos (2 \nu T)-\alpha_2 \delta_1 M-\alpha_2 \delta_{12} M+\alpha_2 M \mu \cos (2 \nu T) \notag \\
b_4 &=-a \alpha_1 \delta_{12} M-a \alpha_1 \delta_2 M+a \alpha_1 M \mu \cos (2 \nu T)-a \alpha_2 \delta_1 M-a \alpha_2 \delta_{12} M\notag\\
&\quad +a \alpha_2 M \mu \cos (2 \nu T)+\delta_1 \delta_{12} M+\delta_1 \delta_2 M-\delta_1 M \mu \cos (2 \nu T)+\delta_{12}^2 M+\delta_{12} \delta_2 M\notag\\
&\quad -2 \delta_{12} M \mu \cos (2 \nu T)-\delta_2 M \mu \cos (2 \nu T)+M \mu^2 \cos ^2(2 \nu T) \notag \\
b_5 &= a \delta_1 \delta_{12} M+a \delta_1 \delta_2 M-a \delta_1 M \mu \cos (2 \nu T)+a \delta_{12}^2 M+a \delta_{12} \delta_2 M-a M (\delta_{12}-\mu \cos (2 \nu T))^2\notag\\
&\quad -2 a \delta_{12} M \mu \cos (2 \nu T)-a \delta_2 M \mu \cos (2 \nu T)+a M \mu^2 \cos ^2(2 \nu T)
\end{align}
\\
For $P_0$ to be a stable fixed point within the linearized analysis, all the eigenvalues must have negative real parts. Since $\lambda_0 = 0$ is a root of the characteristic polynomial, we can consider the remaining eigenvalues by looking at the polynomial $\lambda^5 + b_1\lambda^4 + b_2 \lambda^3 +b_3\lambda^2 + b_4\lambda + b_5$, and from the Routh-Hurwitz criterion, the necessary and sufficient conditions for the roots of this polynomial to have $\hbox{Re} (\lambda_{1,2,3,4,5} <0)$ are:
\begin{align}
\label{crh1}
b_1 &>0 \\ 
\label{crh5}
b_5 &>0 \\ 
\label{crh2}
b_1 b_2 - b_3 &>0 \\ 
\label{crh3}
b_1 (b_2 b_3 + b_5) -b_3^2 - b_1^2 b_4 &>0 \\
\label{chopfcondition}
b_1 (b_2 b_3 b_4 - b_2^2 b_5 + 2 b_4 b_5)-b_3^2 b_4- b_5^2 + b_2b_3b_5-b_1^2b_4^2 &>0 
\end{align}

When the final condition \eqref{chopfcondition} becomes an equality, the characteristic polynomial has one pair of purely imaginary complex conjugate roots. Here we consider the delay parameter $a$ to be the bifurcation parameter. Denote the left hand side of \eqref{chopfcondition} by $f(a)$, which is a fourth degree polynomial in $a$, and whose coefficients, which are too large to include, depend on the remaining parameters. In order to solve the above conditions for parameter regimes which contains a Hopf bifurcation, we fix values for all parameters except $\mu$ and $a$. Then, with our fixed parameter values, we reduce the conditions \eqref{crh1} to \eqref{crh3} along with the condition $f(a) =0$ using computer algebra. The objective is to either obtain conditions on the $\mu$ and $a$ that guarantee a Hopf bifurcation setting with a conjugate pair of imaginary roots, or see that a Hopf bifurcation is not possible for the chosen parameter values. 

In particular we will consider the following parameter set: 
\begin{equation}
\begin{split}
    \alpha_1 = 0.01, \beta_1 = 0.05, \gamma_1 = 3.0, \alpha_2 = 0.01, \beta_2 = 0.05, \gamma_2 = 3.0,\\
    M=0.5, \delta_1 = -0.5, \delta_2 = -0.3, \nu = 2.6, \delta_{12} = 0.3
\end{split}
\end{equation}

Reducing our Routh-Hurwitz Conditions and Hopf Condition for these parameters with computer algebra shows that for no values of $\mu, a,$ and $T$ are all of the conditions satisfied. Thus the system does not have a Hopf bifurcation {\it for the above parameter values}. However, note that a systematic parameter search in Section 5 reveals a rich array of Hopf and other bifurcations, and various dynamical behaviors in our system.


\section{Multiple Scales for the Delayed Landau-Stuart Equation}
In this section, we will use the method of multiple scales to construct analytical approximations for the periodic orbits arising through the Hopf bifurcation of the fixed point of the delayed Landau Stuart system \ref{lsdelay} discussed above. The parameter $a$ will be used as the bifurcation parameter. The limit cycle is determined by expanding about the fixed point using progressively slower time scales. The expansions take the form

\begin{equation}
\label{x1}
x_1 = x_{10} + \sum_{n=1}^3 \delta^n x_{1n}(T_0,T_1,T_2) + ...,
\end{equation}
\begin{equation}
\label{y1}
y_1 = y_{10} + \sum_{n=1}^3 \delta^n y_{1n}(T_0,T_1,T_2) + ...,
\end{equation}
\begin{equation}
\label{z1}
z_1 = z_{10} + \sum_{n=1}^3 \delta^n z_{1n}(T_0,T_1,T_2) + ...,
\end{equation}
\begin{equation}
\label{x2}
x_2 = x_{10} + \sum_{n=1}^3 \delta^n x_{2n}(T_0,T_1,T_2) + ...,
\end{equation}
\begin{equation}
\label{y2}
y_2 = y_{10} + \sum_{n=1}^3 \delta^n y_{2n}(T_0,T_1,T_2) + ...,
\end{equation}
\begin{equation}
\label{z2}
z_2 = z_{10} + \sum_{n=1}^3 \delta^n z_{2n}(T_0,T_1,T_2) + ...,
\end{equation}

\noindent where $T_n = \delta^n t $ and $\delta$ is a small positive non-dimensional parameter that is introduced as a bookkeeping device and will be set to unity in the final analysis. Utilizing the chain rule, the time derivative becomes

\begin{equation}
\frac{d}{dt} = D_0 + \delta D_1 + \delta^2 D_2 + \delta^3 D_3...,
\end{equation}

\noindent where $D_n = \partial / \partial T_n .$ Using the standard expansion for Hopf bifurcations, the delay parameter $a$ is ordered as 

\begin{equation}
\label{aexpand}
a = a_0 + \sum_{n=1}^3 \delta^n a_n(T_0,T_1,T_2) + ...,,
\end{equation}

\noindent where $a_0$ is given by satisfying the Routh-Hurwitz conditions \eqref{rh1} to \eqref{rh4} and \eqref{hopfcondition} with equality. This allows the influence from the nonlinear terms and the control parameter to occur at the same order.\\
Using \eqref{x1}-\eqref{aexpand} in \eqref{lsdelay} and equating like powers of $\delta$ yields equations at $\hbox{O}(\delta^i) ,i =1,2,3$ of the form :

\begin{equation}
\label{L1}
L_1 (x_{1i},y_{1i},z_{1i},x_{2i},y_{2i},z_{2i})= S_{i,1}
\end{equation}
\begin{equation}
\label{L2}
L_2 (x_{1i},y_{1i},z_{1i},x_{2i},y_{2i},z_{2i})= S_{i,2}
\end{equation}
\begin{equation}
\label{L3}
L_3 (x_{1i},y_{1i},z_{1i},x_{2i},y_{2i},z_{2i})= S_{i,3}
\end{equation}
\begin{equation}
\label{L4}
L_4 (x_{1i},y_{1i},z_{1i},x_{2i},y_{2i},z_{2i})= S_{i,4}
\end{equation}
\begin{equation}
\label{L5}
L_5 (x_{1i},y_{1i},z_{1i},x_{2i},y_{2i},z_{2i})= S_{i,5}
\end{equation}
\begin{equation}
\label{L6}
L_6 (x_{1i},y_{1i},z_{1i},x_{2i},y_{2i},z_{2i})= S_{i,6}
\end{equation}

\noindent where the $L_i,i=1,2,3,4,5,6$ are the differential operators

\begin{align}
L_1 (x_{1i},y_{1i},z_{1i},x_{2i},y_{2i},z_{2i}) &= 
D_0x_{1i} + (\varepsilon-1) x_{1i} - \varepsilon x_{3i} + \omega_1 y_{1i} \\
L_2 (x_{1i},y_{1i},z_{1i},x_{2i},y_{2i},z_{2i}) &= 
D_0y_{1i}+(\varepsilon-1) y_{1i}-\varepsilon y_{3i}-\omega_1 x_{1i} \\
L_3 (x_{1i},y_{1i},z_{1i},x_{2i},y_{2i},z_{2i}) &= 
D_0x_{2i} +(\varepsilon-1) x_{2i}-\varepsilon x_{1i}+\omega_2 y_{2i} \\
L_4 (x_{1i},y_{1i},z_{1i},x_{2i},y_{2i},z_{2i}) &= 
D_0y_{2i} +(\varepsilon-1) y_{2i} -\varepsilon y_{1i}-\omega_2 x_{2i}\\
L_5 (x_{1i},y_{1i},z_{1i},x_{2i},y_{2i},z_{2i}) &= 
D_0x_{3i}+a_0 (x_{3i}- x_{2i})\\
L_6 (x_{1i},y_{1i},z_{1i},x_{2i},y_{2i},z_{2i}) &= 
D_0y{3i}+a_0 (y_{3i}- y_{2i})
\end{align}

The source terms $S_{i,j}$ for $i =1,2,3$ and $j = 1,2,3,4,5,6$ i.e. at $\hbox{O}(\delta),\hbox{O}(\delta^2)$, and $\hbox{O}(\delta^3)$ are given as follows. The first order sources $S_{1,j} = 0$ for $j=1,2,3,4,5,6$. The second order sources are:
\begin{align}
\label{2ndsource}
S_{21}&= -D_1x_{11} \notag \\
S_{22}&= -D_1y_{11} \notag \\
S_{23}&= -D_1x_{21} \notag \\
S_{24}&= -D_1y_{21} \notag \\
S_{25}&= - D_1x_{31} + a_1(x_{21}-x_{31})\notag \\
S_{26}&= - D_1y_{31} + a_1(y_{21}-y_{31}) 
\end{align}

and the third order sources are:
\begin{align}
\label{3rdsource}
S_{31}&= -D_2x_{11}-D_1x_{12} -x_{11} y_{11}^2-x_{11}^3\notag \\
S_{32}&= -D_2y_{11}-D_1y_{12} -x_{11}^2 y_{11}-y_{11}^3 \notag \\
S_{33}&= -D_2x_{21}-D_1x_{22} -x_{21} y_{21}^2-x_{21}^3 \notag \\
S_{34}&= -D_2y_{21}-D_1y_{22} -x_{21}^2 y_{21}-y_{21}^3 \notag \\
S_{35}&= -D_2x_{31}-D_1x_{32} +a_1 (x_{22}- x_{32}) + a_2 (x_{21}- x_{31}) \notag \\
S_{3,6}&= -D_2y_{31}-D_1y_{32}+a_1 (y_{22}- y_{32})+a_2 (y_{21}- y_{31})
\end{align}

Next, equation \eqref{L6} may be solved for $y_{2i}$ in terms of $y_{3i}$ . Using this in \eqref{L4}, we can solve for $y_{1i}$ in terms of $y_{3i}$ and $x_{2i}$. Then, we replace $y_{1i}$ in \eqref{L2} and add $\omega_2/\varepsilon$ multiplied by \eqref{L3} to \eqref{L2} which then enables us to solve for $x_{1i}$ in terms of $y_{3i}$. Next, replacing $x_{1i}$ and $y_{1i}$ in equation \eqref{L1}, we can solve for $x_{3i}$ in terms of $y_{3i}$ and $x_{2i}$. Then in \eqref{L5} we can replace $x_{3i}$ and add to it $\omega_1\omega_2/\varepsilon^2$ multiplied by \eqref{L3}, which then allows us to solve for $x_{2i}$ in terms of $y_{3i}$. Finally, using these relations in equation \eqref{L3} gives the composite equation

\begin{equation}
\label{comeq}
L_c w_i=\Gamma_i
\end{equation}

\noindent where 

\begin{equation}
L_c =  D_0^6 + \beta_5 D_0^5 + \beta_4 D_0 ^4 +\beta_3 D_0^3 +\beta_2 D_0^2 +\beta_1 D_0 +\beta_0 
\end{equation}
\noindent and
\begin{align*}
\beta_5 &=  -4 + 2a_0 + 4\varepsilon\\
\beta_4 &=  6+a_0^2+8 a_0 (\varepsilon-1)-12 \varepsilon+6 \varepsilon^2+\omega_1^2+\omega_2^2\\
\beta_3 &= 2 (2 a_0^2 (\varepsilon-1)+(\varepsilon-1) (2-4 \varepsilon+2 \varepsilon^2+\omega_1^2+\omega_2^2)\\
&\quad\quad+a_0 (6-12 \varepsilon+5 \varepsilon^2+\omega_1^2+\omega_2^2)) \\
\beta_2 &= (1-2  \varepsilon+ \varepsilon^2+\omega_1^2) (1-2  \varepsilon+ \varepsilon^2+\omega_2^2)+4 a_0 ( \varepsilon-1) (2-4  \varepsilon+ \varepsilon^2+\omega_1^2+\omega_2^2)\\
&\quad\quad+a_0^2 (6-12  \varepsilon+4  \varepsilon^2+\omega_1^2+\omega_2^2) \\
\beta_1 &= 2 a_0 (-2 \varepsilon^3+(1+\omega_1^2) (1+\omega_2^2)-2 \varepsilon (2+\omega_1^2+\omega_2^2)+\varepsilon^2 (5+\omega_1^2+\omega_1 \omega_2+\omega_2^2)\\
&\quad\quad-a_0 (2+4 \varepsilon^2+\omega_1^2+\omega_2^2-\varepsilon (6+\omega_1^2+\omega_2^2))) \\
\beta_0 &= a_0^2 ((1+\omega_1^2) (1+\omega_2^2)-2 \varepsilon (2+\omega_1^2+\omega_2^2)+\varepsilon^2 (4+\omega_1^2+2 \omega_1 \omega_2+\omega_2^2))
\end{align*}

\noindent The composite source $\Gamma_i$ is equal to 

\begin{align*}
&r_{10} S_{i1} +r_{20} S_{i2} +r_{30} S_{i3} +r_{40} S_{i4} +r_{50} S_{i5} +r_{60} S_{i6} \\
&+r_{11} D_0 S_{i1} +r_{21} D_0S_{i2} +r_{31}D_0 S_{i3} +r_{41}D_0 S_{i4} +r_{51}D_0 S_{i5} +r_{61}D_0 S_{i6} \\
&+r_{12} D_0^2 S_{i1} +r_{22} D_0^2 S_{i2} +r_{32}D_0^2 S_{i3} +r_{42}D_0^2 S_{i4}  +r_{62}D_0^2 S_{i6} \\
&+ r_{23} D_0^3 S_{i2} + r_{33} D_0^3 S_{i3} + r_{43} D_0^3 S_{i4} + r_{63} D_0^3 S_{i6}\\
&- a_0 D_0^4 S_{i4} + (4-a_0-4\varepsilon) D_0^4 S_{i6}- D_0^5 S_{i6}
\end{align*}

\noindent where

\begin{align*}
r_{10} &= -a_0^2 (\varepsilon-1) \varepsilon (\omega_1+\omega_2) \\
r_{20} &= a_0^2 \varepsilon (-1+2 \varepsilon+\omega_1 \omega_2) \\
r_{30} &= -a_0^2 (-2 \varepsilon \omega_2+(1+\omega_1^2) \omega_2+\varepsilon^2 (\omega_1+\omega_2)) \\
r_{40} &= a_0^2 (1+2 \varepsilon^2+\omega_1^2-\varepsilon (3+\omega_1^2)) \\
r_{50} &= -a_0 (\varepsilon-1) \varepsilon^2 (\omega_1+\omega_2) \\
r_{60} &= a_0 (2 \varepsilon^3-(1+\omega_1^2) (1+\omega_2^2)+2 \varepsilon (2+\omega_1^2+\omega_2^2)-\varepsilon^2 (5+\omega_1^2+ \omega_1 \omega_2 + \omega_2^2)) \\
r_{11} &= -a_0 \varepsilon (\varepsilon-1+a_0) (\omega_1+\omega_2) \\
r_{21} &= a_0 \varepsilon (-1-2 a_0 (\varepsilon-1)+2 \varepsilon-\varepsilon^2+\omega_1 \omega_2) \\
r_{31} &= -a_0 (1+2 a_0 (\varepsilon-1)-2 \varepsilon+\varepsilon^2+\omega_1^2) \omega_2 \\
r_{41} &= -a_0 ((\varepsilon-1) (1-2 \varepsilon+\varepsilon^2+\omega_1^2)+a_0 (3-6 \varepsilon+2 \varepsilon^2+\omega_1^2)) \\
r_{51} &= -a_0\varepsilon^2(\omega_1 + \omega_2) \\
r_{61} &= -(1-2 \varepsilon+\varepsilon^2+\omega_1^2) (1-2 \varepsilon+\varepsilon^2+\omega_2^2)-2 a_0 (\varepsilon-1) (2-4 \varepsilon+\varepsilon^2+\omega_1^2+\omega_2^2)
\end{align*}
\begin{align*}
r_{12} &= -a_0\varepsilon(\omega_1+\omega_2) \\
r_{22} &= -a_0 \varepsilon (-2+a_0+2 \varepsilon) \\
r_{32} &= -a_0 (-2+a_0+2 \varepsilon) \omega_2 \\
r_{42} &= -a_0 (3+3 a_0 (\varepsilon-1)-6 \varepsilon+3 \varepsilon^2+\omega_1^2) \\
r_{62} &= -2 (\varepsilon-1) (2-4 \varepsilon+2 \varepsilon^2+\omega_1^2+\omega_2^2)-a_0 (6-12 \varepsilon+5 \varepsilon^2+\omega_1^2+\omega_2^2) \\
r_{23} &= -a_0\varepsilon \\
r_{33} &= -a_0\omega_2 \\
r_{43} &= -a_0 (-3+a_0+3 \varepsilon) \\
r_{63} &= -6-4 a_0 (\varepsilon-1)+12 \varepsilon-6 \varepsilon^2-\omega_1^2-\omega_2^2 \\
\end{align*}

We use \eqref{comeq} later to identify and suppress secular terms in the solutions of \eqref{L1}-\eqref{L6}\\
Let us now turn to finding the solutions of \eqref{L1}-\eqref{L6}, solving order by order in the usual way. \\
For $i=1$ or $\hbox{O}(\epsilon)$ we know $S_{1,k}= 0$ for $k=1,...,6$. Hence we pick up a solution for the first order fields using the eigenvalues (from the previous section) at Hopf bifurcation, which we denote $\lambda_1 = i\omega$ and it's complex conjugate $\lambda_2$, i.e.

\begin{equation}
\label{y31}
y_{31}=\alpha[T_1,T_2,T_3]e^{-i\omega t} + \beta[T_1,T_2,T_3]e^{i\omega t} 
\end{equation}

\noindent
where $\beta = \bar{\alpha}$ is the complex conjugate of $\alpha$ since $\lambda_2= \bar{\lambda}_1$ and $ y_{31}$ is real. As is evident, the $\alpha$ and $\beta$ modes correspond to the center manifold where $\lambda_{1,2}$ are purely imaginary and where the Hopf bifurcation occurs. Since we wish to construct and analyze the stability of the periodic orbits which lie in the center manifold, we suppress the other eigenvalues with non-zero real parts. 

Using \eqref{y31} in \eqref{L1}-\eqref{L6} for $i=1$ and the process used to derive the composite equation we have:

\begin{equation}
\label{y21}
y_{21} = \frac{e^{-i \omega T_0}}{a_0} \big((a_0-i \omega) \alpha[T1,T2,T3]+e^{2 i \omega T_0} (a_0+i \omega) \beta[T1,T2,T3]\big)
\end{equation}

\begin{align}
\label{x21}
x_{21} &= \frac{e^{-i \omega T_0}}{a_0 (\omega_1+\omega_2) \left(a_0 \left(\varepsilon^2+\omega_1 \omega_2\right)-(\varepsilon-1) \omega_1 \omega_2\right)} \bigg(i \omega^3 \left(a_0^2+6 a_0 (\varepsilon-1)+3 \varepsilon^2\right.\notag\\
&\quad-6 \left.\varepsilon+\omega_1^2+\omega_2^2+3\right) \left(\alpha(T_1,T_2,T_3)-e^{2 i \omega T_0} \beta(T_1,T_2,T_3)\right)\notag\\
&\quad-\omega^2 \left(3 a_0^2 (\varepsilon-1)+a_0 \left(5 \varepsilon^2-12 \varepsilon+2 \omega_1^2+\omega_1 \omega_2+2 \omega_2^2+6\right)\right.\notag\\
&\quad\left.+(\varepsilon-1) \left(\varepsilon^2-2 \varepsilon+\omega_1^2-\omega_1 \omega_2+\omega_2^2+1\right)\right) \left(\alpha(T_1,T_2,T_3)+e^{2 i \omega T_0} \beta(T_1,T_2,T_3)\right)\notag\\
&\quad-i \omega \left(a_0^2 \left(2 \varepsilon^2-6 \varepsilon+\omega_1^2+\omega_1 \omega_2+\omega_2^2+3\right)+a_0 (\varepsilon-1) \left(\varepsilon^2-4 \varepsilon+2 \left(\omega_1^2+\omega_2^2+1\right)\right)\right.\notag\\
&\quad\left.+\omega_1 \omega_2 \left(-\varepsilon^2+2 \varepsilon+\omega_1 \omega_2-1\right)\right) \left(\alpha(T_1,T_2,T_3)-e^{2 i \omega T_0} \beta(T_1,T_2,T_3)\right)\notag\\
&\quad-a_0 \left(a_0 \left(2 \varepsilon^2-\varepsilon \left(\omega_1^2+\omega_1 \omega_2+\omega_2^2+3\right)+\omega_1^2+\omega_1 \omega_2+\omega_2^2+1\right)\right.\notag\\
&\quad\left.-\omega_1 \omega_2 (2 \varepsilon+\omega_1 \omega_2-1)\right) \left(\alpha(T_1,T_2,T_3)+e^{2 i \omega T_0} \beta(T_1,T_2,T_3)\right)\notag\\
&\quad+\omega^4 (2 a_0+3 \varepsilon-3) \left(\alpha(T_1,T_2,T_3)+e^{2 i \omega T_0} \beta(T_1,T_2,T_3)\right)-i \omega^5 \alpha(T_1,T_2,T_3)\notag\\
&\quad+i \omega^5 e^{2 i \omega T_0} \beta(T_1,T_2,T_3)\bigg)
\end{align}

\begin{align}
\label{x11}
x_{11} &= \frac{e^{-i \omega T_0}}{a_0 \varepsilon (\omega_1+\omega_2)} \bigg(\left(a_0 \left(\varepsilon (-2-2 i \omega)-\omega^2+2 i \omega+\omega_2^2+1\right)\right.\notag\\
&\quad\left.-i \omega \left(\varepsilon^2+\varepsilon (-2-2 i \omega)-\omega^2+2 i \omega+\omega_2^2+1\right)\right) \alpha(T_1,T_2,T_3)\notag\\
&\quad+e^{2 i \omega T_0} \left(a_0 \left(2 i \varepsilon (\omega+i)-\omega^2-2 i \omega+\omega_2^2+1\right)+i \omega \left(\varepsilon^2+2 i \varepsilon (\omega+i)\right.\right.\notag\\
&\quad\left.\left.-\omega^2-2 i \omega+\omega_2^2+1\right)\right) \beta(T_1,T_2,T_3)\bigg)
\end{align}

\noindent
where we have omitted $y_{11}$ and $x_{31}$ as the expressions for them are too long to include. Now that the first order solutions are known, the second-order sources $S_{21},S_{22},S_{23},S_{24},S_{25},S_{26}$ may be evaluated using \eqref{2ndsource}. Computing the second-order composite source $\Gamma_2$, we find that the entire source is secular and that the Setting the coefficients of the secular $e^{\pm i\omega t}$ terms in these sources to zero yields

\begin{equation}
\label{1result}
D_1 \alpha = \frac{\partial \alpha}{\partial T_1} = 0, D_1 \beta = \frac{\partial \beta}{\partial T_1} = 0
\end{equation}

Next, using the second-order sources, and \eqref{1result} , the second-order particular solution is taken
in the usual form to balance the zeroth and second harmonic terms at this order, i.e., 

\begin{equation}
\label{y32}
y_{32}=y_{32,0}+y_{32,2}e^{2i\omega t}
\end{equation}

Then since the entire second order source was secular, upon removing the secular terms with \eqref{1result} we find the second order source is now zero. Thus using \eqref{y32} in \eqref{comeq} for $i=2$ we find the coefficients in the second-order particular solution are $y_{32,0} = y_{32,2} = 0$, thus $y_{32} = 0$. Then using $y_{32}$ in \eqref{L1}-\eqref{L6} for $i=2$, together with the second-order sources, yields that the other second-order fields are also zero, 

$$y_{12}=y_{22} = x_{12}=x_{22} = x_{32} = 0$$

Using these, together with the first-order results, we may evaluate the coefficients of the secular terms in the composite source $\Gamma_3$, from \eqref{3rdsource} and \eqref{comeq}. Suppressing these secular, first-harmonic, terms to obtain uniform expansions yields the final equation for the evolution of the coefficients in the linear solutions on the slow second-order time scales

\begin{equation}
\label{nform}
\frac{\partial \beta}{\partial T_2} = C_1 \beta + C_2 \alpha \beta ^2 
\end{equation}

\noindent where the very large expressions for the coefficients $C_i$ are omitted for the sake of brevity. 

This equation \eqref{nform} is the normal form, or simplified system in the center-manifold, in the vicinity of the Hopf bifurcation point. We shall now proceed to compare the predictions for the post-bifurcation dynamics from this normal form with actual numerical simulations.


\section{Numerical Results and Discussion}

\subsection{Landau-Stuart Equation}

We may immediately make two additional points here regarding the Hopf bifurcation. In most systems \citep{Choud}, the Hopf bifurcation may occur either below or above the critical value of the system's chosen bifurcation parameter, and one needs to test which in fact occurs. Since we have chosen the delay $a$ as bifurcation parameter, and larger delays or lower $a$ values have a stabilizing effect, we know that for our delayed Landau-Stuart system, the post-Hopf regime is for $a$ values larger than the $a_{Hopf}$ value found using the second root of the polynomial in the last equation of Section 3.1. For $a < a_{Hopf}$, the strong delay stabilizes the oscillations and yields a stable fixed point. This is thus the regime of Amplitude Death(AD) for the system caused by the delay. The  $a = a_{Hopf}$ point is thus the exact value of the delay parameter where AD sets in, and this may be precisely pinpointed here via the semi-analytic treatment in Section 3.1. 

Note also that, in principle, the Hopf bifurcation might be either supercritical with stable oscillations seen above $a = a_{Hopf}$ or at weaker delays, or subcritical where the Hopf-created periodic orbit is unstable and coexists with the stable fixed point in the $a < a_{Hopf}$ or Amplitude Death regime. In the latter case, there would be no nearby system attractor for $a > a_{Hopf}$, and the dynamics in that regime would feature any of the three following scenarios:
a. jumping to a distant periodic attractor if one exists, b. flying off to infinity in finite time (an attractor at infinity), or c. an aperiodic attractor on which the system orbits evolve. 

However, we may plausibly rule out the occurrence of this latter, subcritical Hopf scenario. This is because the undelayed Landau-Stuart system is a robust oscillator showing stable periodic behavior, that, under the effect of delay, persists in the $a > a_{Hopf}$ regime of a post-supercritical Hopf bifurcation, while being reduced to Amplitude Death by stronger delays for $a < a_{Hopf}$.
This does in fact turn out to be correct, as will be verified below via both the normal form and numerical simulations.

By approximating the flow of the system in a computer model, we can easily analyze the behavior of the system for various sets of parameters. Here we will consider the case in section 3.1 where $\varepsilon = 2, \omega_1 = 15$, and $\omega_2 = 15$ and values of $a$ around the Hopf bifurcation value $a_{Hopf} \approx 5.63185$.

\begin{figure}[H]
\centering
\centerline{\includegraphics[width=0.5\textwidth]{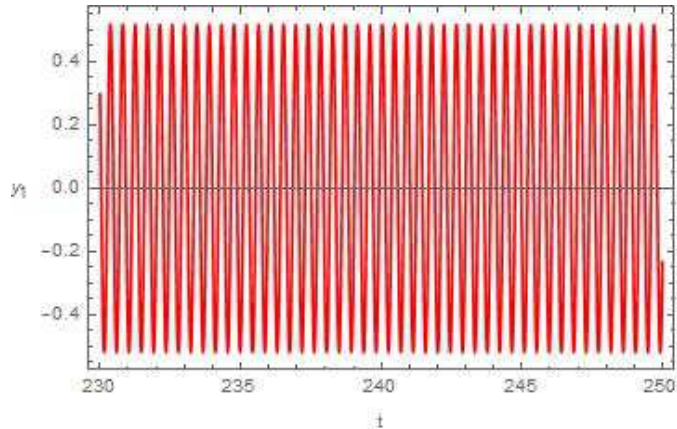}}
\caption{\label{fig:lsy1a10}Periodic oscillations in $y_1$ for $a = 10$.}
\end{figure}

\begin{figure}[H]
\centering
\centerline{\includegraphics[width=0.5\textwidth]{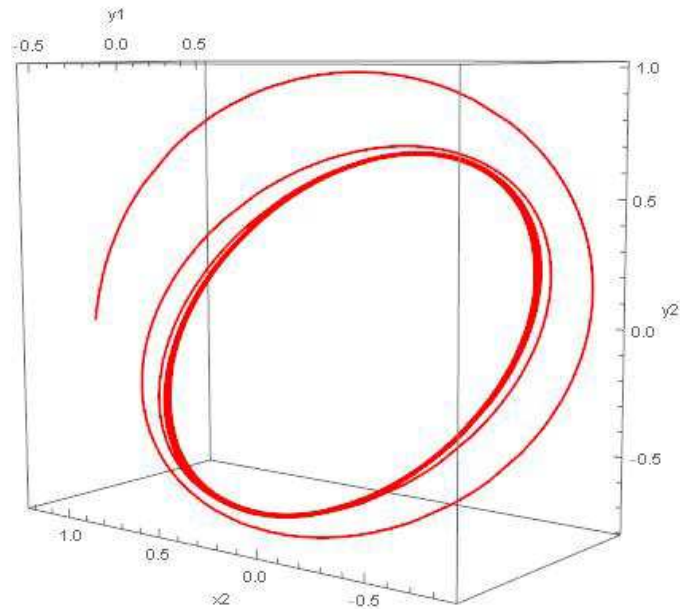}}
\caption{\label{fig:lsphasea10} The limit cycle in $(x_2,y_1,y_2)$ phase space for the parameters of Figure \ref{fig:lsy1a10} and the approach from the initial conditions. }
\end{figure}

\begin{figure}[H]
\centering
\centerline{\includegraphics[width=0.5\textwidth]{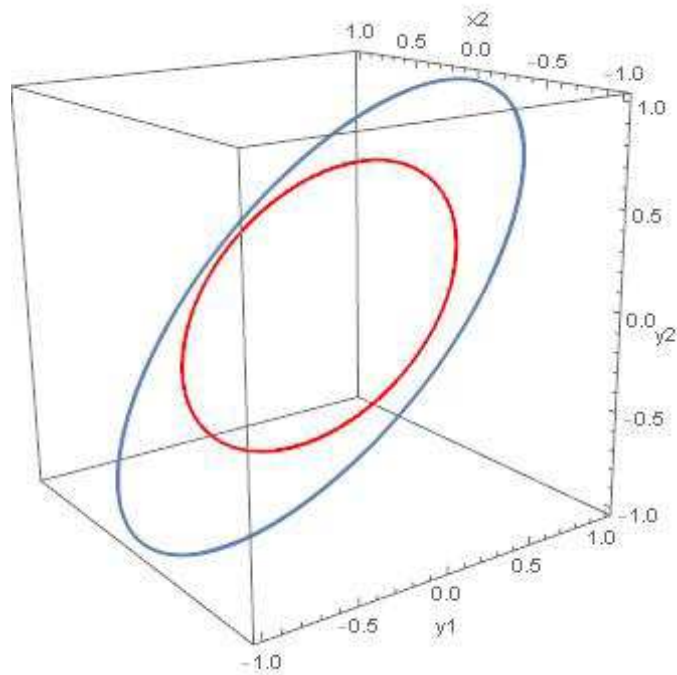}}
\caption{\label{fig:lscomparea10} The smaller delayed limit cycle in red and undelayed limit cycle in blue plotted in $(x_2,y_1,y_2)$ phase space for the parameters of Figure \ref{fig:lsy1a10}. }
\end{figure}

Figures \ref{fig:lsy1a10} through \ref{fig:lscomparea10} show the limit cycle for $a = 10$ above the Hopf bifurcation value $a_{Hopf}$. As predicted from the normal form, and our plausibility argument above, we have stable periodic behavior above the bifurcation point as shown in Figure \ref{fig:lsy1a10} for $y_1(t)$. Figure 2 shows the limit cycle in $(x_2,y_1,y_2)$ phase space and the approach from the initial conditions. Figure 3 shows both the delayed (in red) and undelayed (in blue) limit cycles in $(x_2,y_1,y_2)$ phase space from which we can see the stabilizing effect of the delay causing the limit cycle to shrink towards the fixed point at the origin, as well as rotate in phase space.  

\begin{figure}[H]
\centering
\centerline{\includegraphics[width=0.5\textwidth]{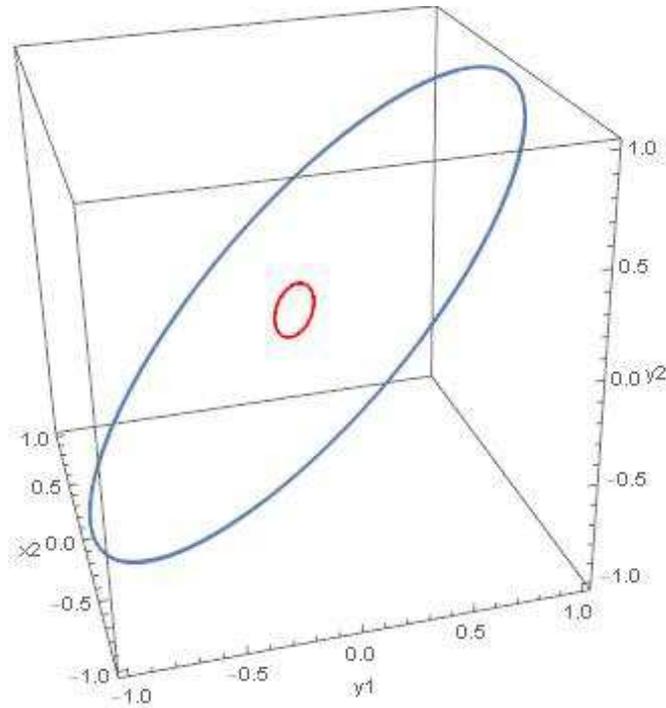}}
\caption{\label{fig:lscomparea5-73} The delayed limit cycle in red and undelayed limit cycle in blue plotted in $(x_2,y_1,y_2)$ phase space for $a = 5.73$. }
\end{figure}

Figure \ref{fig:lscomparea5-73} shows the limit cycle for $a = 5.73$ just above the bifurcation point $a_{Hopf}$ in red and the undelayed system in blue in $(x_2, y_1,y_2)$ phase space. Here we can see that, as we further decrease the parameter $a$ towards the bifurcation value or increase the delay, the limit cycle continues to shrink towards the fixed point at the origin. 

\begin{figure}[H]
\centering
\centerline{\includegraphics[width=0.5\textwidth]{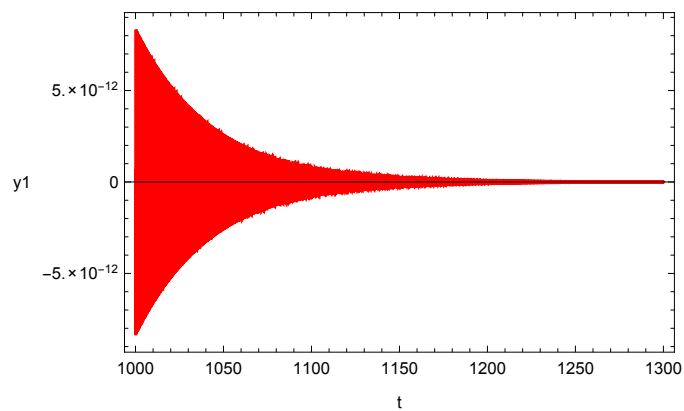}}
\caption{\label{fig:lsy1a5-4} Amplitude death in $y_1$ for $a = 5.4$. }
\end{figure}

\begin{figure}[H]
\centering
\centerline{\includegraphics[width=0.5\textwidth]{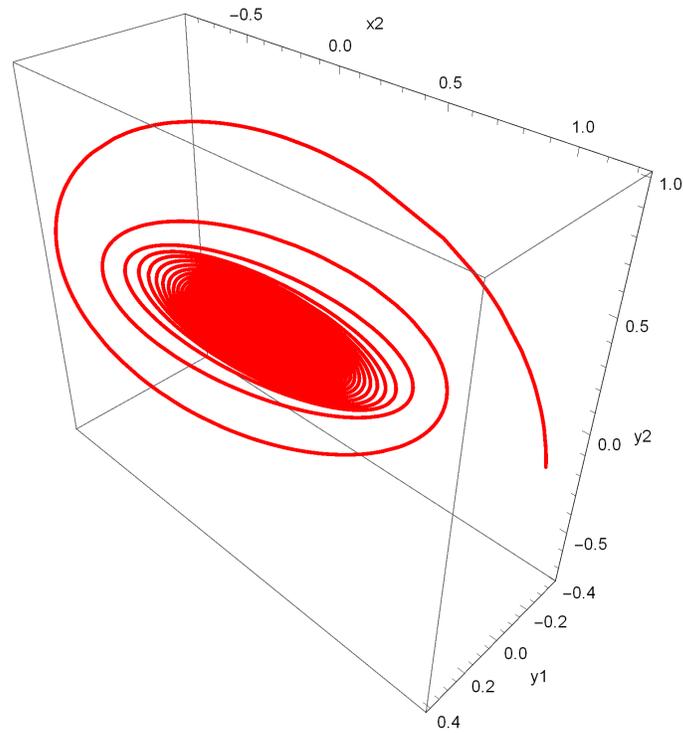}}
\caption{\label{fig:lscomparea5-4} The delayed limit cycle in red tending to the origin and undelayed limit cycle in blue plotted in $(x_2,y_1,y_2)$ phase space for $a = 5.4$. }
\end{figure}

Next, Figures  \ref{fig:lsy1a5-4} and  \ref{fig:lscomparea5-4} show the delayed solution for an even larger delay $a=5.4$ which is now below the bifurcation value $a_{Hopf}$. Here, we see the system exhibit Amplitude Death as the solutions spiral towards the now stabilized origin. 

\begin{figure}[H]
\centering
\centerline{\includegraphics[width=0.6\textwidth]{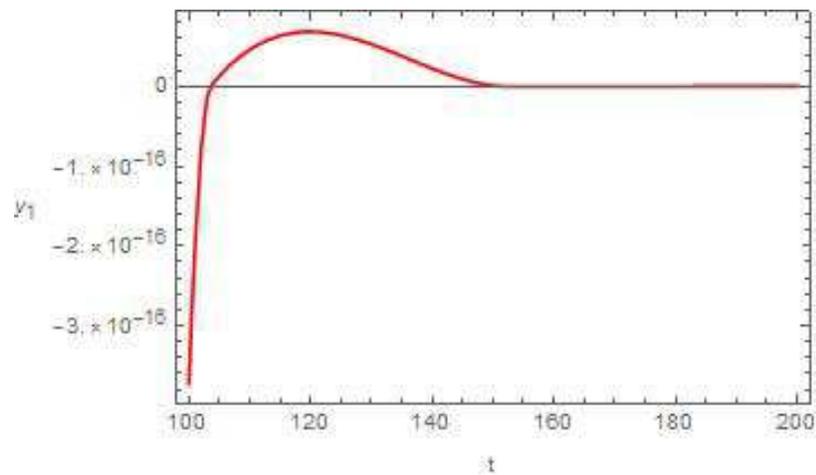}}
\caption{\label{fig:lsy1a2} Amplitude death in $y_1$ for $a = 2$. }
\end{figure}

\begin{figure}[H]
\centering
\centerline{\includegraphics[width=0.5\textwidth]{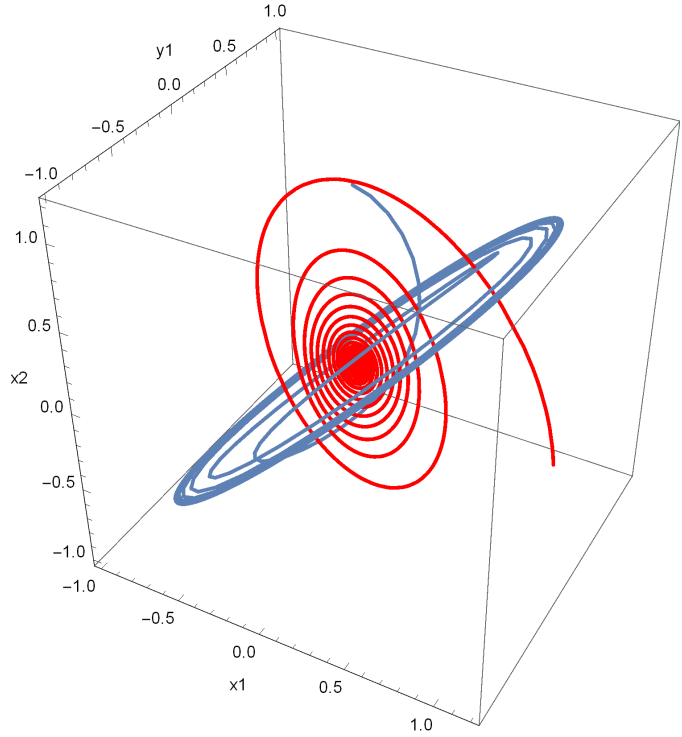}}
\caption{\label{fig:lscomparea2} The delayed limit cycle in red tending to the origin and undelayed limit cycle in blue plotted in $(x_2,y_1,y_2)$ phase space for $a = 2$. }
\end{figure}

Finally Figure \ref{fig:lsy1a2} shows the delayed time series for $y_1$ when $a = 2$.  Figure \ref{fig:lscomparea2} shows both the delayed solution in red and the undelayed solution in blue, as well as their approach from the initial conditions, where  the delayed system again exhibits Amplitude Death. We also observe that the smaller the value of $a$, or the greater the delay, the faster the approach to the origin. 

In this delayed system, as mentioned above, the limit cycles in the $a > a_{Hopf}$ regime are very robust, as one might expect since the undelayed Landau-Stuart system is well-known to demonstrate stable periodic behavior over wide ranges of the system parameters. However, it is quite possible that these robust limit cycles might be quickly disrupted by secondary symmetry breaking, cyclic-fold, flip, transcritical, or Neimark-Sacker bifurcations  when some  other system parameter is changed. To investigate this, for chosen values of $a$ well above $a_{Hopf}$, we varied the other system parameters deep into this post-Hopf regime, i.e. far from the starting values $\varepsilon = 2, \omega_1 = 15$, and $\omega_2 = 15$ used above. The post-supercritical Hopf limit cycle proves extremely robust under variation of all three of these parameters. No further complex dynamics arises in this delayed system from additional bifurcations of the Hopf-created limit cycles, not surprisingly since the undelayed Landau-Stuart system is a stable oscillator over a wide range of these parameters.

\subsection{Chaotic System}
Since our preliminary search for a Hopf bifurcation yielded a negative result for one set of parameters, let us first vary the value of the delay parameter $a$ and study its effect on the system. While the effect of delay can be predicted to be stabilizing, a much more complex set of dynamical behaviors occurs for this case, including a rich array of evolving system attractors as $a$, as well as other system parameters, are varied. Hence, the latter part of this sub-section will also systematically consider the bifurcations and dynamics as the other important parameter $\mu$, which measures the strength of the parametric excitation, is varied. This will systematically reveal a variety of dynamical behaviors.

\subsubsection{Chaotic Case $\mu = 0.5$}
\begin{figure}[H]
\centering
\centerline{\includegraphics[width=0.5\textwidth]{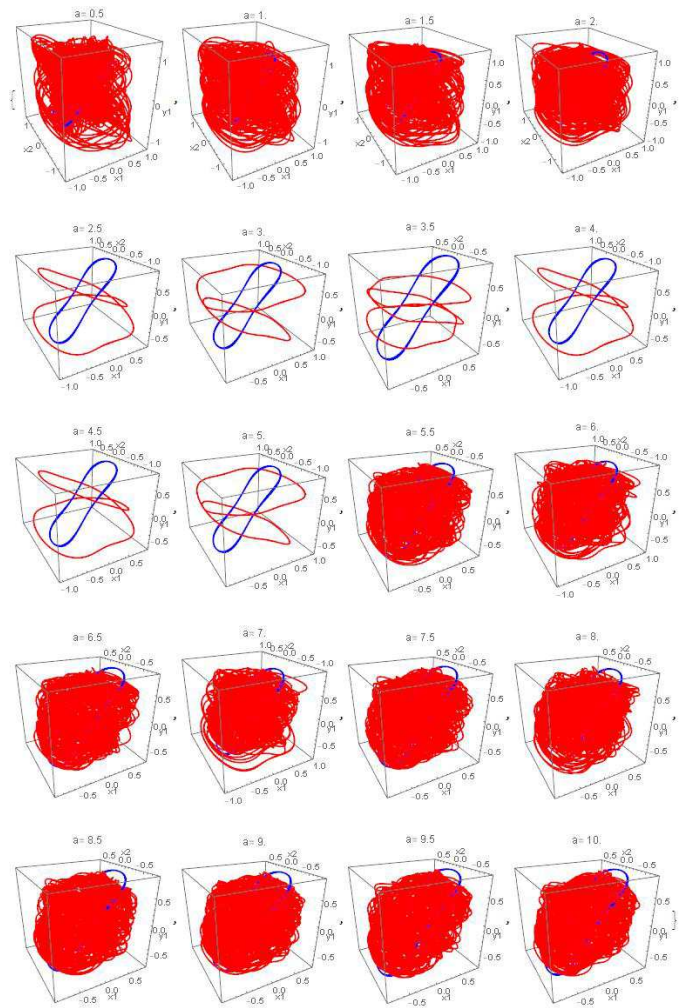}}
\caption{\label{fig:c1-chaotic-unforced-transition}The delayed (red) and undelayed (blue) solutions of the system in the chaotic case ($\mu = 0.5$) with no forcing ($q=0$) for various values of the delay parameter $a$. }
\end{figure}

Figure \ref{fig:c1-chaotic-unforced-transition} show solutions in $(x_1,x_2,y_1)$ phase space of the delayed attractor in red and undelayed attractor in blue in the chaotic case of $\mu = 0.5$ (having one positive Lyapunov exponent, three negative exponents, and a fifth one along the time coordinate and hence always having value zero). We first consider the system in the absence of forcing ($q=0$) as values of the delay parameter $a$ range from $a=0.5$ to $a=10$. Here we observe 3 types of behavior as we vary $a$, the first being a cocoon shaped structure surrounding the undelayed attractor which occurs for $a=0.5$ to $a = 2$, $a=5.5$ to $a=10$. The second type of behavior is a double loop type structure for the delayed solutions, again surrounding the undelayed attractor, and occurring in two different ways, the first oriented as for $a= 3$ and the second oriented as in the case $a=4$ (a rotated version of $a =3$). The final type of behavior is the case $a = 3.5$ where we see a slightly more complicated looping structure surrounding the undelayed attractor.


\begin{figure}[H]
\centering
\centerline{\includegraphics[width=0.5\textwidth]{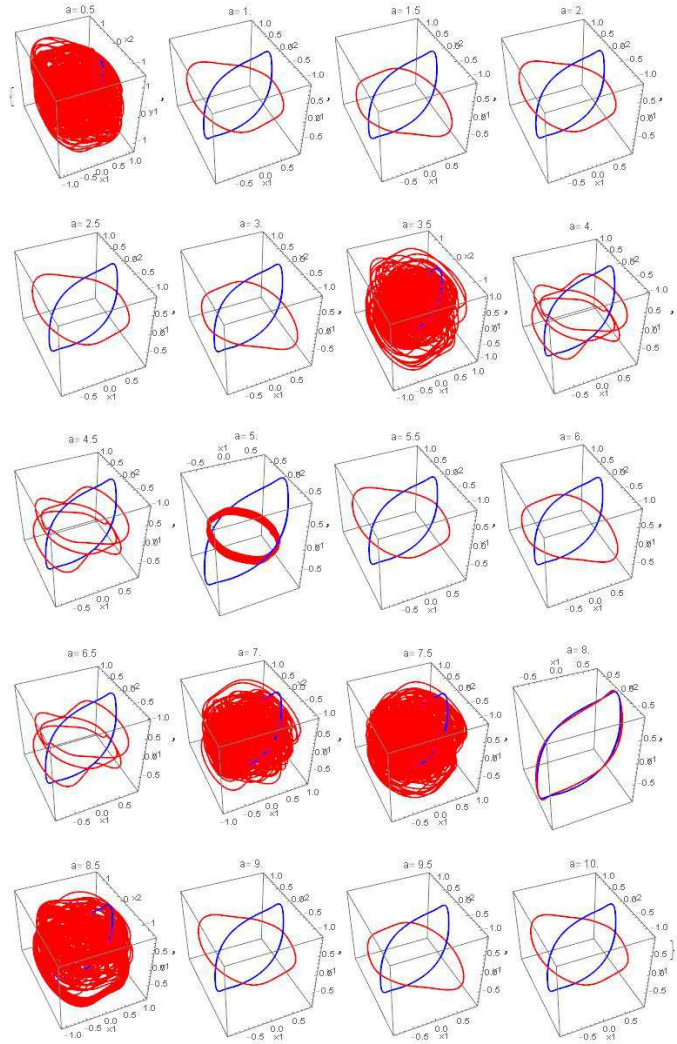}}
\caption{\label{fig:c1-chaotic-transition-forced-0-5}The delayed (red) and undelayed (blue) solutions of the system in the chaotic case ($\mu = 0.5$) with forcing ($q=0.5$) for various values of the delay parameter $a$.}
\end{figure}

Next, in Figure \ref{fig:c1-chaotic-transition-forced-0-5} we have plots in $(x_1,x_2,y_1)$ phase space of the delayed attractor in red and undelayed attractor in blue in a forced chaotic case with $\mu = 0.5$ and $q = 0.5$ as we vary the delay parameter. As in the unforced case, for several values of $a$, the delayed solution is like a cocoon around the undelayed attractor. For the cases $a=1$ to $a=3$, $a=5.5, 6$, and $a=9$ to $a=10$ we see the delayed solution is now a thin horizontal loop around the undelayed attractor. For the cases $a=4, 4.5, 6.5$ the delay makes the shape of the attractor much more complicated with several loops now surrounding the undelayed attractor. In the case $a=5$ we see the delay results in a much thicker smaller attractor while in the case $a=8$ we see the delayed attractor is very similar to the undelayed case. Both are expected results, with the stabilizing effect of the smaller $a$ or larger delay shrinking the attractor, while the case with larger $a$ has only weak delay and so does not differ appreciably from the undelayed system.

\begin{figure}[H]
\centering
\centerline{\includegraphics[width=0.5\textwidth]{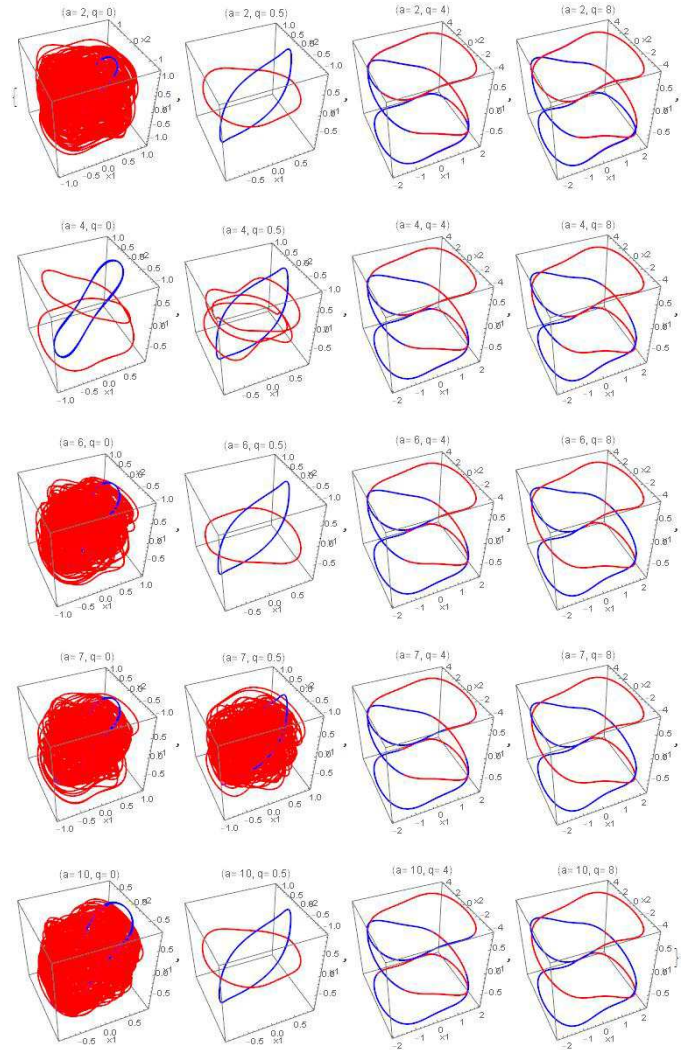}}
\caption{\label{fig:c1-chaotic-delay-forcing-array}The delayed (red) and undelayed (blue) solutions of the system in the chaotic case ($\mu = 0.5$) for values of $a=2,4,6,7,10$ and $q= 0,0.5,4,8$}
\end{figure}

Finally in Figure \ref{fig:c1-chaotic-delay-forcing-array} we have solutions of the of the delayed and undelayed system for $\mu = 0.5$ as we vary both the delay parameter $a$ (increasing down the columns) and forcing parameter $q$ (increasing down the rows). The first thing to observe is that the most varied behavior occurs in the unforced case, and that as we increase the forcing the effect of the delay decreases. For instance, for $q=4,8$ the undelayed and delayed systems have very similar solutions even as we vary the delay strength. Again this is intuitively something one would expect, with the increasing $q$ or forcing having a destabilizing effect that counteracts the stabilizing effect of increasing delay as $a$ is reduced.

Note that, unlike in the case of the delayed Landau-Stuart system, even for very large delays or small values of $a$ the system does not exhibit complete Amplitude Death or stabilization of the chaotic behavior to either a stable limit cycle or, even further, to a stable fixed point. As we shall see below, transition from chaotic regimes to synchronized periodic oscillations on limit cycles (sometimes referred to as Oscillation Death, or perhaps more accurately Chaos Death in this case) is indeed possible if we look more widely in our parameter space.

\subsubsection{Hyperchaotic Case $\mu = 2$}
In this section we look at numerically generated solutions of the system \eqref{c1delaysys} for hyperchaotic cases with $\mu = 2$ (having two positive, two negative, and one zero (along the time coordinate) Lyapunov exponent).

\begin{figure}[H]
\centering
\centerline{\includegraphics[width=0.65\textwidth]{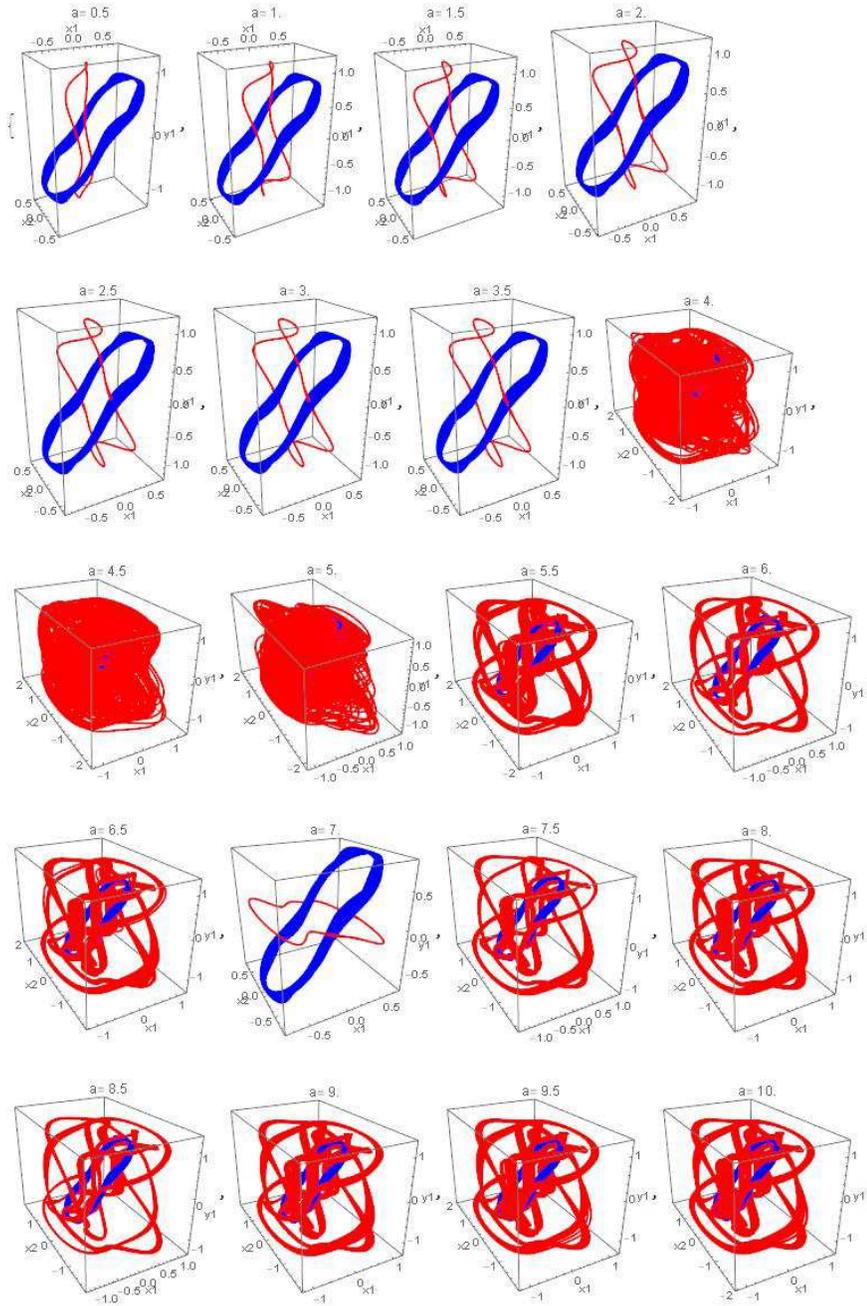}}
\caption{\label{fig:c1-hc-unforced-transition}The delayed (red) and undelayed (blue) solutions of the system in the hyperchaotic case ($\mu = 2$) with no forcing ($q=0$) for various values of the delay parameter $a$.}
\end{figure}

Figure \ref{fig:c1-hc-unforced-transition} shows plots in $(x_1,x_2,y_1)$ phase space of the delayed attractor in red and undelayed attractor in blue in the hyperchaotic case $\mu = 2$ with no forcing ($q= 0$) as values of the delay parameter $a$ range from $a=0.5$ to $a =10$. We see that the delayed attractor is initially thin and long, and oriented vertically. As we increase $a$ from $a=0.5$ to $a=3.5$ the top and bottom ends of the attractor form a loop. From $a=4$ to $a=5$ we see the attractor does not have a more amorphous shape, forming a cocoon around the undelayed attractor. For $a=5.5$ through $a=10$, the delay causes the system's attractor to take on a much more complicated shape that loops around the undelayed attractor, with the exception of $a=7$ where the delayed solution forms a horizontal loop around the delayed attractor instead. 

\begin{figure}[H]
\centering
\centerline{\includegraphics[width=0.5\textwidth]{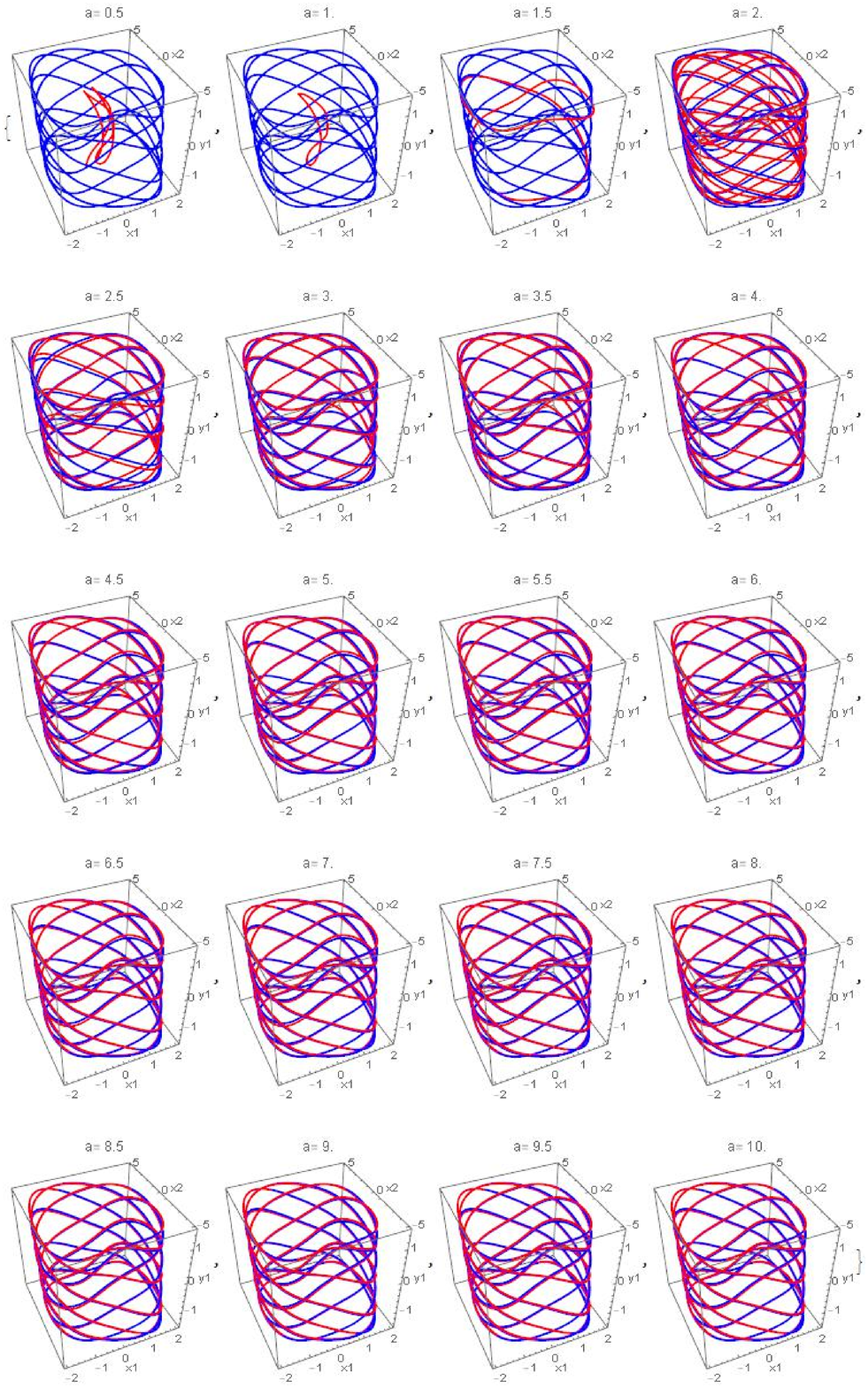}}
\caption{\label{fig:c1-hc-forced-2-5}The delayed (red) and undelayed (blue) solutions of the system in the hyperchaotic case ($\mu = 2$) with forcing ($q=2.5$) for various values of the delay parameter $a$.}
\end{figure}

Next in Figure \ref{fig:c1-hc-forced-2-5} we have plots in $(x_1,x_2,y_1)$ phase space of the delayed attractor in red and undelayed attractor in blue in the forced hyperchaotic case, $\mu = 2$ and $q = 2.5$ as we vary the delay parameter. From this figure we see that at higher values of $a$ or weak delay, the delayed and undelayed solutions are, as one would expect, almost the same. At small values of $a$, the stabilizing effect of the stronger delay causes the attractor to become much smaller than for the undelayed case. Since the destabilizing effect of the forcing is quite strong for $q=2.5$, note that only strong delay (corresponding to when $a$ is small) has a significant effect on the system attractor. 

\begin{figure}[H]
\centering
\centerline{\includegraphics[width=0.5\textwidth]{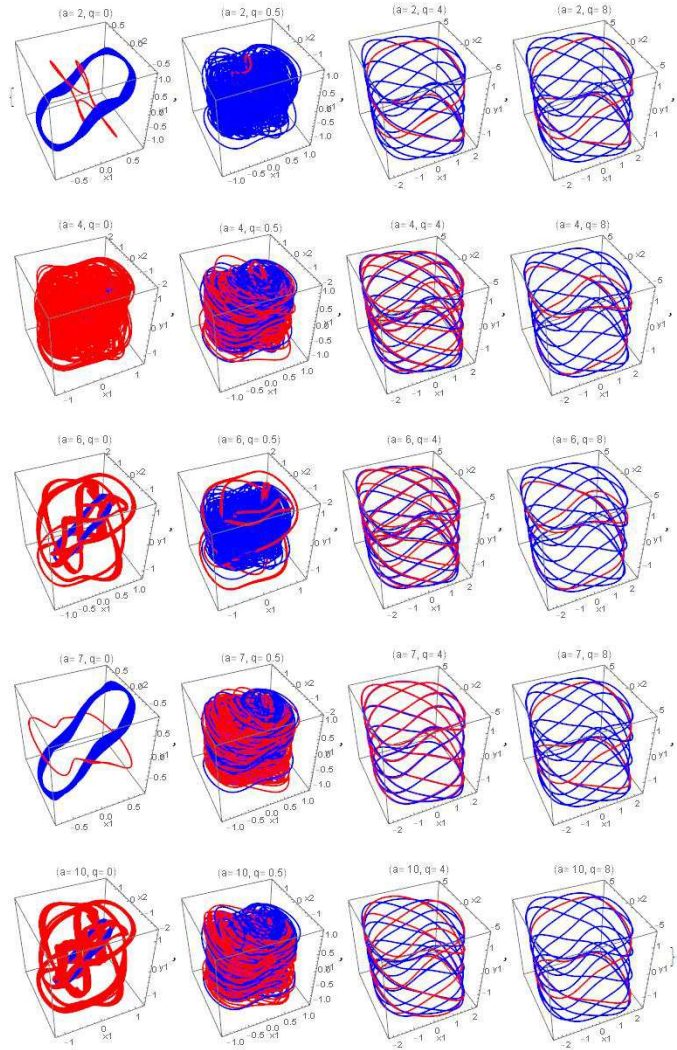}}
\caption{\label{fig:c1-hc-delay-forcing-array}The delayed (red) and un delayed (blue) solutions of the system in the hyperchaotic case ($\mu = 2$) for values of $a=2,4,6,7,10$ and $q= 0, 0.5,4,8$}
\end{figure}

In Figure \ref{fig:c1-hc-delay-forcing-array} we have solutions of the of the delayed and undelayed system as we vary both the delay parameter $a$ (increasing down the columns) and forcing parameter $q$ (increasing down the rows). We see that for no forcing the introduction of the delay causes very different behavior as the delay strength varies as we saw in Figure \ref{fig:c1-hc-unforced-transition}. However, increasing the forcing parameter we see that the effects of the delay for different values of $a$ become similar. We also see that at the higher forcing value $q=8$ the delayed orbits are simpler than the undelayed orbit. In particular the case $q=8$ shows that unlike in Figure \ref{fig:c1-hc-forced-2-5} it is not always the case that the delay only has significant effects on the system at smaller values of $a$. This is again expected, as the very strong destabilizing effect of this large forcing would be partially counteracted even by weak delays.

\subsection{Varying the Parametric Forcing}

The above gives a general idea about the effects of the delay and forcing on the system dynamics. In order to understand the various possible dynamical regimes, and the transitions between them,
more comprehensively, we shall next consider the effect of systematically increasing the other, and perhaps most important, system parameter $\mu$ which controls the parametric forcing.

We consider the case of weak delay with $a = 10$, although smaller $a$ values show qualitatively similar behavior. 
At small $\mu 0.1$, we see periodic dynamics, as seen in the phase plot of Figure 15, and the power spectral density
of Figure 16 which shows a single narrow peak at $\omega \simeq 0.137$.

\begin{figure}[H]
\centering
\centerline{\includegraphics[width=0.5\textwidth]{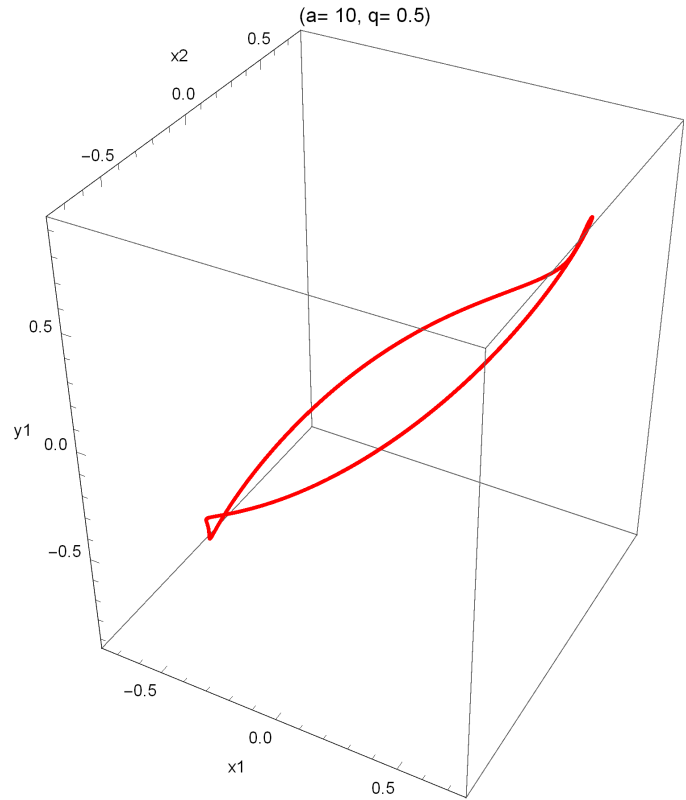}}
\caption{\label{fig:figure15}The phase space plot for $\mu = 0.1$, and $a=10, q = 0.5$.}
\end{figure}

\begin{figure}[H]
\centering
\centerline{\includegraphics[width=0.5\textwidth]{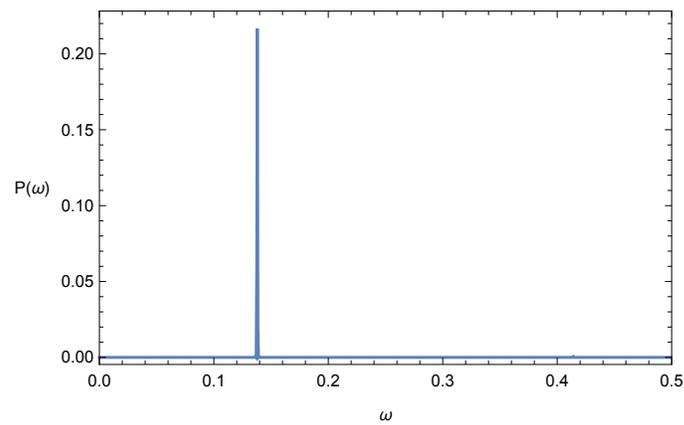}}
\caption{\label{fig:figure16}The power spectral density for $\mu = 0.1$, and $a=10, q = 0.5$.}
\end{figure}

There is a complete cascade of period doublings for $\mu \in (0.1, 0.11)$, leading to a more complex chaotic attractor with one positive Lyapunov exponent at $\mu = 0.11$, as seen in the phase plot of Figure 17, and the broad features in the power spectral density of Figure 18.

\begin{figure}[H]
\centering
\centerline{\includegraphics[width=0.5\textwidth]{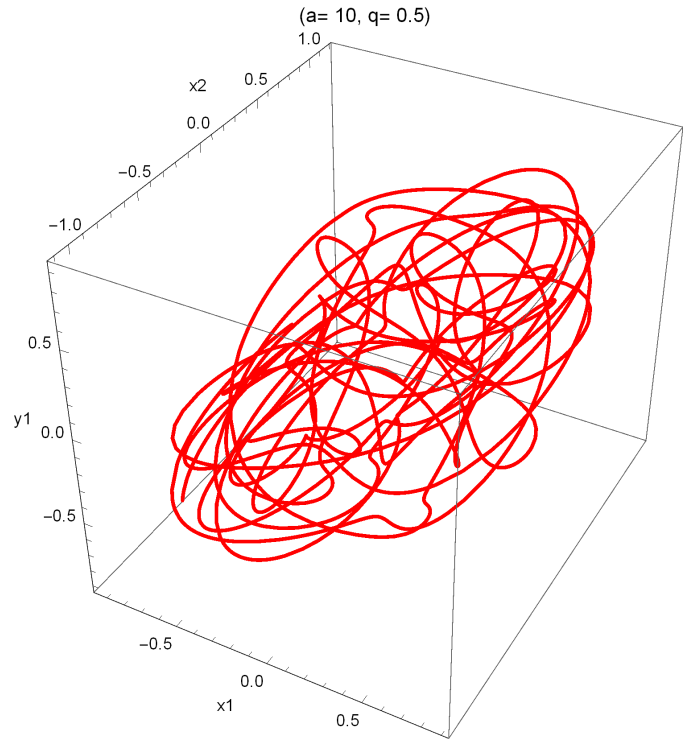}}
\caption{\label{fig:figure17}The phase space plot for $\mu = 0.11$, and $a=10, q = 0.5$.}
\end{figure}

\begin{figure}[H]
\centering
\centerline{\includegraphics[width=0.5\textwidth]{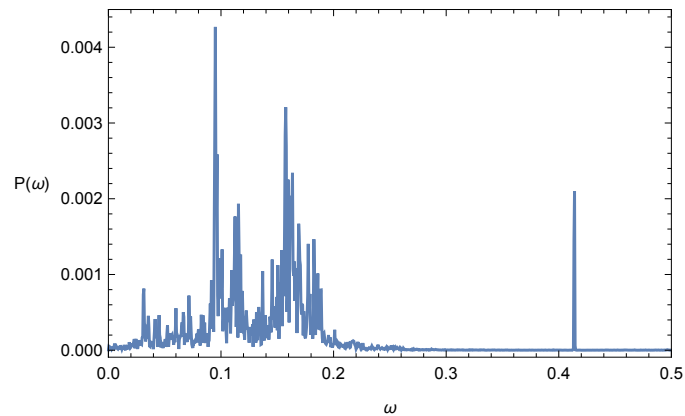}}
\caption{\label{fig:figure18}The broad chaotic features in the power spectral density for $\mu = 0.11$, and $a=10, q = 0.5$. Note the secondary single peak at $\omega \simeq 0.416$.}
\end{figure}

The chaotic behavior persists over the window $\mu \in (0.11, 3.43)$ and then is destroyed in a boundary crisis for $\mu \in (3.43, 3.44)$, leading into a new period doubled attractor at $\mu = 3.44$ with a dominant single peak at $\omega \simeq 0.208$ as seen in Figures 19 and 20. This corresponds to a synchronized state of the two oscillators.

\begin{figure}[H]
\centering
\centerline{\includegraphics[width=0.5\textwidth]{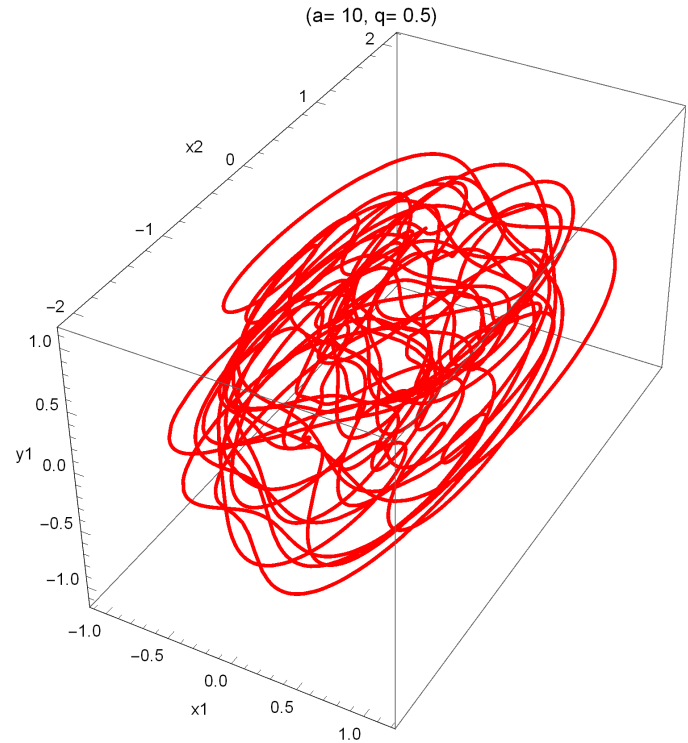}}
\caption{\label{fig:figure19}The phase space plot for $\mu = 3.44$, and $a=10, q = 0.5$.}
\end{figure}

\begin{figure}[H]
\centering
\centerline{\includegraphics[width=0.5\textwidth]{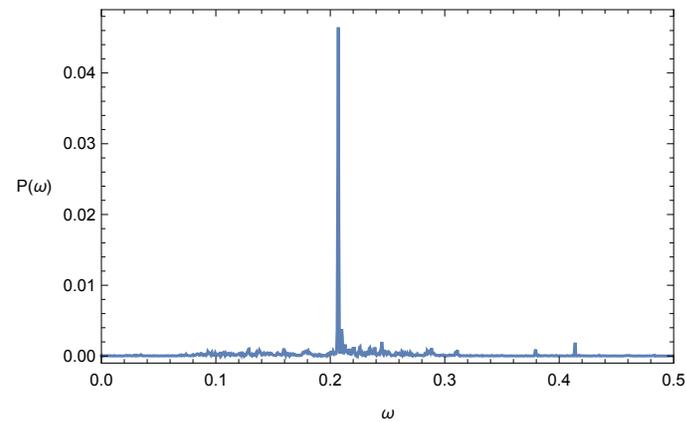}}
\caption{\label{fig:figure20}The single peaked power spectral density for $\mu = 0.11$, and $a=10, q = 0.5$, with $\omega \simeq 0.208$ and a very small secondary peak still persisting at $\omega \simeq 0.416$.}
\end{figure}

This periodic attractor then immediately undergoes a symmetry breaking bifurcation for $\mu \in (3.44, 3.45)$, as shown in the power spectral density plot of Figure 21 where the symmetry breaking gives rise to the peak at the second harmonic frequency of $\omega \simeq 0.416$ 

\begin{figure}[H]
\centering
\centerline{\includegraphics[width=0.5\textwidth]{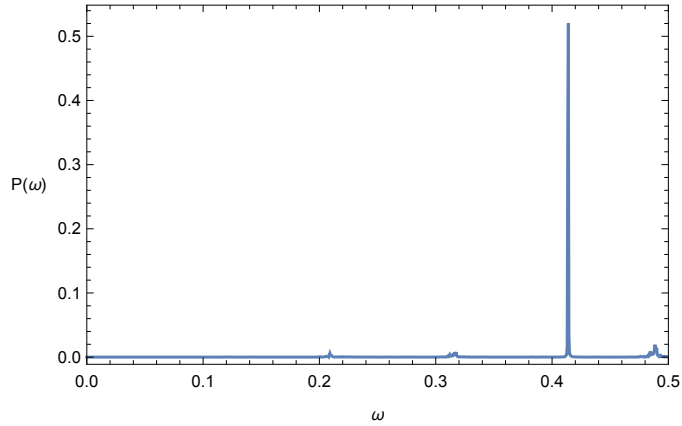}}
\caption{\label{fig:figure22}The single peaked power spectral density for $\mu = 3.45$, and $a=10, q = 0.5$ with $\omega \simeq 0.416$, the second harmonic of the frequency in Figure 20.}
\end{figure}

As $\mu$ is increased further, a small secondary peak at $\omega \simeq 0.24$ is created as the oscillators losing synchronization near $\mu  \simeq 5.3$. The behavior is thus now two-period quasiperiodic, and this persists till $mu = 83.41$, as seen in Figures 22 and 23, showing the attractor and the double-peaked power spectrum at that value.

\begin{figure}[H]
\centering
\centerline{\includegraphics[width=0.5\textwidth]{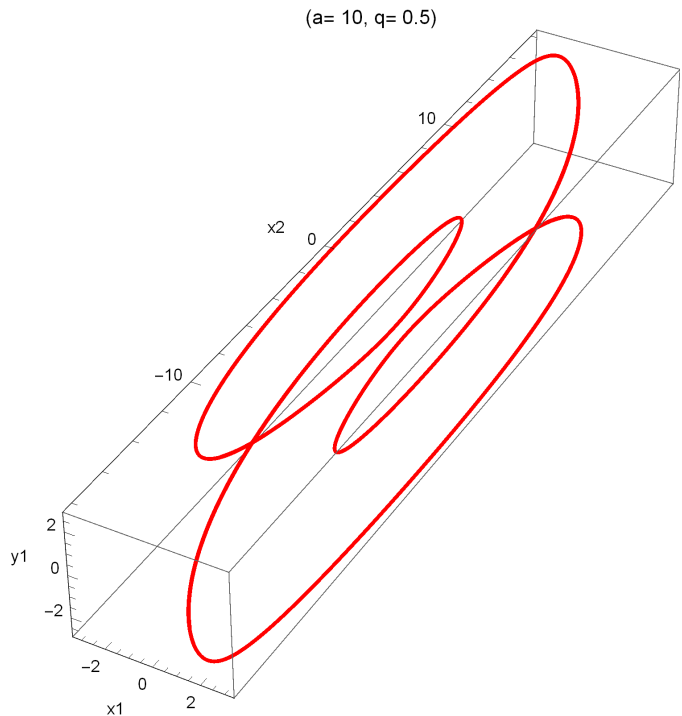}}
\caption{\label{fig:figure23}The two-period quasiperiodic attractor for $\mu = 83.41$, and $a=10, q = 0.5$.}
\end{figure}

\begin{figure}[H]
\centering
\centerline{\includegraphics[width=0.5\textwidth]{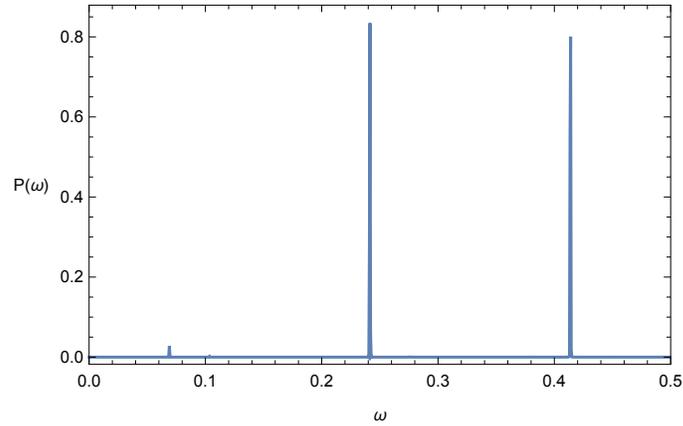}}
\caption{\label{fig:figure24}The power spectral density for $\mu = 83.41$, and $a=10, q = 0.5$, with $\omega \simeq 0.208$ and a second peak at an incommensurate frequency $\omega \simeq 0.24$.}
\end{figure}

Following this, there is a cascade of torus doublings for $\mu \in (83.4113, 83.4114)$, leading to a more complex chaotic attractor at $\mu = 83.42$ with one positive Lyapunov exponent, as seen in the phase space plot of Figure 24, and the broad features in the power spectral density of Figure 25.

\begin{figure}[H]
\centering
\centerline{\includegraphics[width=0.5\textwidth]{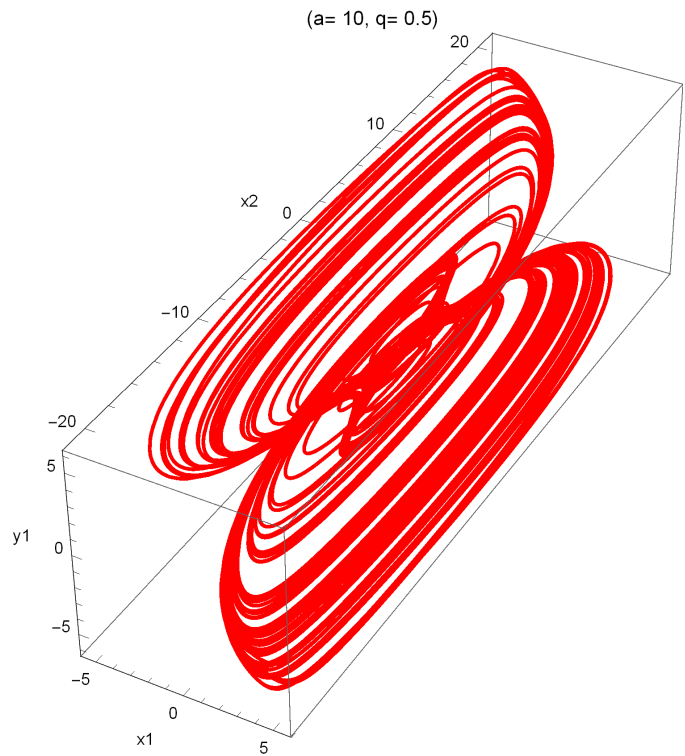}}
\caption{\label{fig:figure25}The phase space plot for $\mu = 83.42$, and $a=10, q = 0.5$ after a sequence of torus doublings.}
\end{figure}

\begin{figure}[H]
\centering
\centerline{\includegraphics[width=0.5\textwidth]{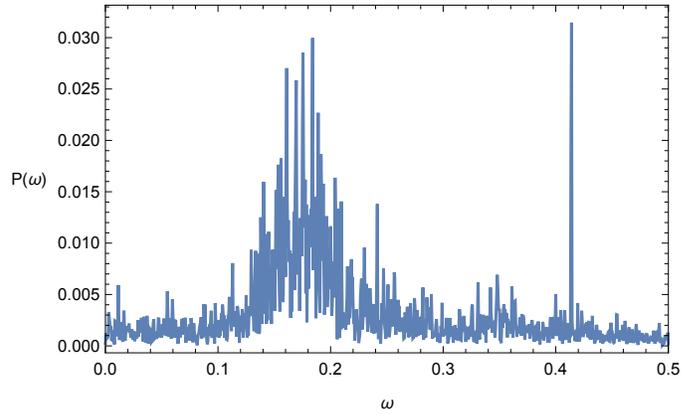}}
\caption{\label{fig:figure26}The broad chaotic features in the power spectral density for $\mu = 83.42$, and $a=10, q = 0.5$. Note the secondary single peak at $\omega \simeq 0.416$.}
\end{figure}

As $\mu$ in raised further, the chaotic attractor is destroyed by a boundary crisis at $\mu \simeq 83.45$ as seen in Figures 26 and 27. In the latter, the earlier two peaks in the power spectral density persist, but sidebands and a new peak at $\omega \simeq 0.095$ have been created.

\begin{figure}[H]
\centering
\centerline{\includegraphics[width=0.5\textwidth]{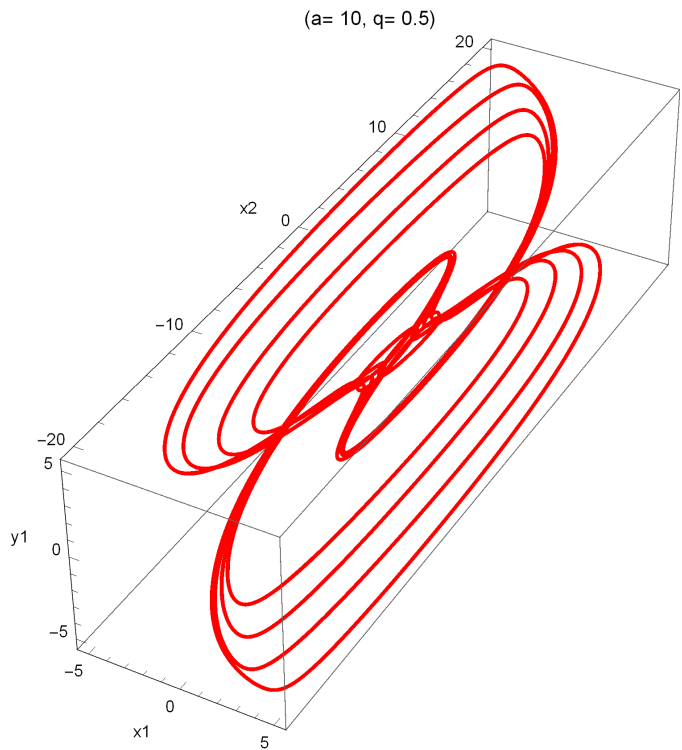}}
\caption{\label{fig:figure27}The phase space plot for $\mu = 83.45$, and $a=10, q = 0.5$.}
\end{figure}

\begin{figure}[H]
\centering
\centerline{\includegraphics[width=0.5\textwidth]{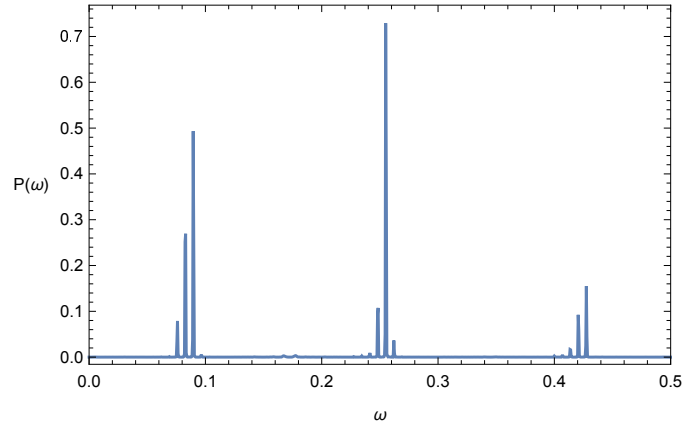}}
\caption{\label{fig:figure28}The power spectral density for $\mu = 83.45$, and $a=10, q = 0.5$. The earlier two peaks in the power spectral density persist, but sidebands and a new peak at $\omega \simeq 0.095$ have been created.} 
\end{figure}

Finally this exterior crisis begins to terminate in a stable quasiperiodic attractor at $\mu \simeq 83.48$ as seen in Figure 28 where the earlier two peaks in the power spectral density persist, but a new peak at $\omega \simeq 0.175$ has been created.

\begin{figure}[H]
\centering
\centerline{\includegraphics[width=0.5\textwidth]{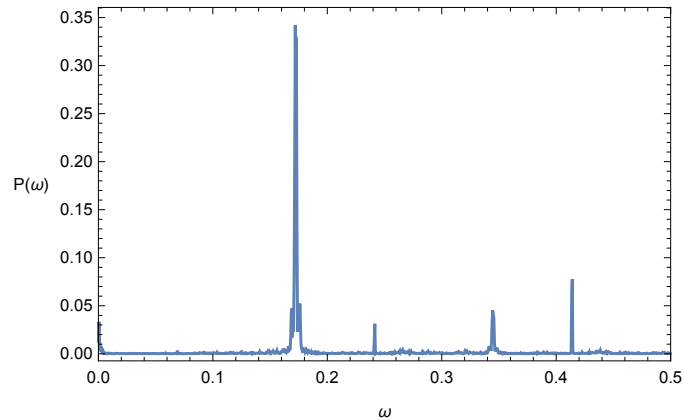}}
\caption{\label{fig:figure30}The power spectral density for $\mu = 83.48$, and $a=10, q = 0.5$. The earlier two peaks in the power spectral density persist, but a new peak at $\omega \simeq 0.175$ has been created.} 
\end{figure}

For slightly higher $\mu \simeq 83.5$, a new second harmonic peak is born at $\omega \simeq 0.35$ by symmetry breaking, and the crisis terminates with the cleaner-looking power spectrum at $\mu \simeq 85$ seen in Figure 29.

\begin{figure}[H]
\centering
\centerline{\includegraphics[width=0.5\textwidth]{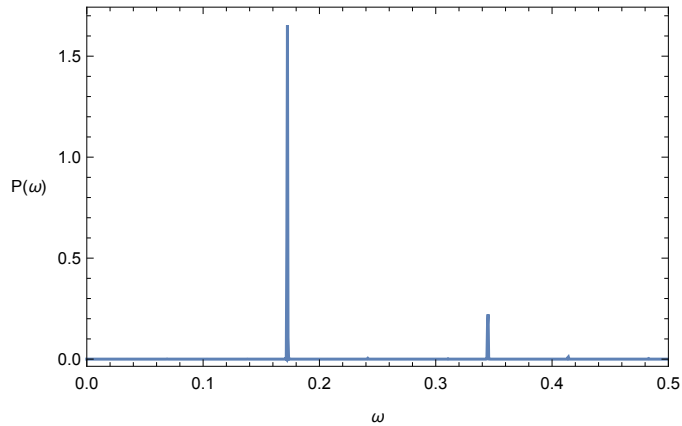}}
\caption{\label{fig:figure31}The power spectral density for $\mu = 85$, and $a=10, q = 0.5$. The new peak at $\omega \simeq 0.175$  and its second harmonic now remain.} 
\end{figure}

We shall end our bifurcation sequence here for this case, as the general features are clear by now.

To conclude our numerical results, let us very briefly consider the case of strong delay with $a = 0.1, q = 8$, where we use a stronger forcing to partly balance the stabilizing effect of the very large delay. Now the range of periodic behavior with $\omega \simeq 0.416$ at low values of $\mu$ persists up to $\mu \simeq 94.63$ after which a second frequency $\omega \simeq 0.24$ comes in via Hopf bifurcation. Further bifurcations and changes in system dynamics as $\mu$ is raised then mimic those discussed above for the weak delay case, except that they occur at significantly larger values of $\mu$.

\section{Results and Conclusions}

We have comprehensively analyzed the effects of distributed 'weak generic kernel' delays on the coupled Landau-Stuart system, as well as a chaotic oscillator system with parametric forcing. As expected, increasing the delay by reducing the delay parameter $a$ is stabilizing, with its Hopf bifurcation value (dependent, of course, on the other system parameters) being a point of exact Amplitude death for both the Landau-Stuart and the chaotic van der Pol-Rayleigh parametrically forced system. In the Landau-Stuart system, the Hopf-generated limit cycles for $a > a_{Hopf}$  are very robust under large variations of all other system parameters beyond the Hopf bifurcation point, and do not undergo
further symmetry breaking, cyclic-fold, flip, transcritical or Neimark-Sacker bifurcations. This is to be expected as the corresponding undelayed systems are robust oscillators over very wide ranges of their respective parameters. 

Numerical simulations reveal strong distortion and rotation of the limit cycles in phase space as the parameters are pushed far into the post-Hopf regime, and also enable tracking of other features, such as how the oscillation amplitudes and time periods of the physical variables on the limit cycle attractor change as the delay and other parameters are varied. For the chaotic system, very strong delays may still lead to the onset of AD (even for relatively large values of the system forcing which tends to oppose this stabilization phenomenon). 

Varying of the other important system parameter, the parametric excitation, leads to a rich sequence of evolving dynamical regimes, with the bifurcations leading from one into the next being carefully tracked numerically here.

\end{document}